\documentclass[aps,prd,reprint,groupedaddress,floatfix]{revtex4-2}
\usepackage{natbib}
\usepackage{graphicx}	
\usepackage{amsmath}	
\usepackage{amssymb}	
\usepackage{tabularx} 
\usepackage{amsmath,bm}  
\usepackage{amssymb} 
\usepackage{graphicx} 
\usepackage[margin=1in,letterpaper]{geometry} 
\usepackage[final]{hyperref} 
\hypersetup{
	colorlinks=true,       
	linkcolor=blue,        
	citecolor=blue,        
	filecolor=magenta,     
	urlcolor=blue         
}

\usepackage{color}
\usepackage{xcolor}
\usepackage{soul}

\begin{document}

\title{Probing dark matter with polarimetry techniques}

\author{A.~Ejlli}
\email{EjlliA@cardiff.ac.uk}
\affiliation{Gravity Exploration Institute, Cardiff University, Cardiff CF24 3AA, United Kingdom}	
\author{S.M.~Vermeulen}
\affiliation{Gravity Exploration Institute, Cardiff University, Cardiff CF24 3AA, United Kingdom}
\author{E.~Schwartz}
\affiliation{Gravity Exploration Institute, Cardiff University, Cardiff CF24 3AA, United Kingdom}
\author{L.~Aiello}
\affiliation{Gravity Exploration Institute, Cardiff University, Cardiff CF24 3AA, United Kingdom}
\author{H.~Grote}
\affiliation{Gravity Exploration Institute, Cardiff University, Cardiff CF24 3AA, United Kingdom}

\raggedbottom

\begin{abstract}

\noindent In this work, we propose polarimetry experiments to search for low-mass (sub-eV) bosonic field dark matter, including axions and axion-like particles. We show that a polarimetry configuration consisting of a thick birefringent solid inside a Fabry-Pérot cavity is exceptionally sensitive to scalar field dark matter, which may cause oscillatory variations in the solid's thickness and refractive index. In addition, we show that a reconfiguration of this polarimetry experiment, in which two quarter-wave plates are placed inside the Fabry-Pérot cavity instead of a thick birefringent solid, is very sensitive to axion-like particles. We investigate the possibility of using cross-correlation of twin polarimeters to increase the sensitivity of the experiment, which in turn could allow us to explore unexplored parts of the parameter space and potentially detect a signal in either dark matter scenario.

\end{abstract}
\maketitle

\section{Introduction}
The $\Lambda$CDM (Lambda-cold dark matter) model \cite{aghanim2020planck}, also known as the Standard Model of Cosmology, is a six-parameter model that fits all available data with a high degree of precision and concisely summarises our current knowledge of the history and composition of the universe. According to the $\Lambda$CDM model, the majority of the present Universe consists of a combination of dark energy (68\%), dark matter (DM) (27\%), and baryons (5\%), such as nucleons. The many DM theories and possible interactions with the Standard Model make for a diffuse but active search effort, as DM's unknown Nature remains one of the most pressing issues in modern physics. 

Weakly interacting massive particles (WIMPs) were the most promising candidates for DM during the course of the previous few decades. However, a number of very sensitive detectors \cite{aprile2017xenon1t, akerib2013large,zhang2019dark}, have yet to detect such particles and future upgrades will be limited by the solar neutrino background \cite{baudis2014neutrino}. In addition, the Large Hadron Collider has set stringent limits on super-symmetry, the theoretical underpinning for massive particles with weak interactions \cite{canepa2019searches}. 

In recent years, the concept of searching for sub-eV-mass DM candidates has attracted considerable attention. Numerous well-motivated candidates of this sort exist, including the canonical axion, axion-like particles (ALPs), and dilatons, which may all manifest as a coherently oscillating classical field. In this paper, we concentrate on scalar field DM~\cite{stadnik2015searching, stadnik2015can} and pseudoscalar axion-like particles~\cite{PhysRevD.88.035023, Ringwald_2014, Ringwald_2014, farina2017photo, PhysRevD.104.062008}. The search for scalar-field DM, axion-like particles, and quantum gravity are examples of recent revolutionary concepts in laser interferometry's direct application to physics that goes beyond the gravitational field \cite{grote_novel_2019, vermeulen2021direct, aiello_constraints_2021, vermeulen2021experiment, martynov_quantum-enhanced_2020}. 

This work presents a way to utilize polarimetry with high sensitivity in order to identify potential couplings of scalar field DM and axion-like particles. In 1979, E.~Iacopini and E.~Zavattini published seminal work with the objective of determining how to detect vacuum magnetic birefringence \cite{euler_uber_1935, erber_velocity_1961} with polarimetry \cite{iacopini_experimental_1979}. Unlike Michelson interferometry, which looks for the relative displacement of two orthogonal arms, polarimetry is sensitive to the relative phase variation of two orthogonal polarization components. Today, the scheme is still used for measuring minute birefringence \cite{zavattini2022polarimetry}, and it provides the highest sensitivity for measuring the vacuum magnetic birefringence \cite{ejlli_pvlas_2020}. 

In the polarimetry setup proposed in this paper, an oscillating scalar or pseudo-scalar DM field is expected to produce a relative phase modulation between the two orthogonal polarizations at the same frequency as the field oscillation, which could be detected. Low-mass bosonic dark matter is assumed to have a long coherence length (relative to the measurement apparatus' dimensions) \cite{derevianko_detecting_2018}, such that two identical polarimeters close together would measure the same signal and may be cross-correlated in a search for these dark-matter fields.

The paper structure is as follows:  In section \ref{sec:theory}, we recapitulate how scalar DM may cause thickness changes in highly birefringent materials, and how the axion-like field interacts with polarized light. These two distinct coupling mechanisms would both produce a signal in a polarimeter setup. In section~\ref{sec:method}, we investigate these two scenarios by analytically calculating the dark matter-induced phase difference in a polarimeter with a Fabry-Pérot cavity. We also evaluate the expected polarimeter noise in section~\ref{sec:noise}. In section \ref{sec:Prospects}, we present projections of the experiment's integrated sensitivity for both cross-correlation of twin polarimeters and a single polarimeter. Lastly, in section~\ref{sec:discussion}, we summarise our findings and discuss DM searches using polarimetry in the context of other direct searches.

\section{Theory}
\label{sec:theory}
\subsection{Scalar Field DM}
A light scalar field with a mass (${10^{-21} < m_\phi < 10}$~eV) created in the early universe would manifest in the present day as an oscillating classical field \cite{derevianko_detecting_2018,arvanitaki_searching_2015}:

\begin{equation}\label{eq:DMfield}
    \phi(t,\vec{r})=\phi_0\,\cos(\omega_{\phi}t-\vec{k}_{\phi}\cdot\vec{r}),
\end{equation}

where $\omega_{\phi}=(m_{\phi}c^2)/\hbar$ is the angular Compton frequency, $\vec{k}_{\phi}=(m_{\phi}\vec{v}_{\mathrm{obs}})/\hbar$ the wave vector and $\vec{v}_{\mathrm{obs}}$ is the velocity relative to the observer, and the amplitude $\phi_0=(\hbar\sqrt{2\rho_{\mathrm{local}}})/(m_{\phi}c)$, where $\rho_{\mathrm{local}}$ is the local DM density. 
If other components of dark matter are present, the amplitude of the oscillatory field could be much smaller \cite{PhysRevD.103.083535, XENON:2020fgj}. We thus assume that the local dark matter density is fully made up of the undiscovered field.

Fields like this could couple to the electromagnetic part of the Standard Model; the simplest such couplings are linear in $\phi$ \cite{antypas_new_2022,damour_phenomenology_2010}:

\begin{equation}\label{eq:lagrangian}
    \mathcal{L}_{\mathrm{int}}=\frac{\phi}{\Lambda_{\gamma}}\frac{F_{\mu\nu}F^{\mu\nu}}{4}-\frac{\phi}{\Lambda_e}m_e\bar{\psi}_e\psi_e, 
\end{equation}

where $F_{\mu\nu}$ is the electromagnetic field tensor, $\psi_e$ is the electron wave function, and $\Lambda_{\gamma}$ and $\Lambda_e$ are the electromagnetic and electronic coupling constants, respectively.
These additional terms entail corrections to electromagnetic interactions that are not present in the Standard Model. The corrections can be effected by considering the electron mass and the fine structure constant to be coupled to the scalar field

\begin{equation}
\alpha' = \alpha\left(1+\frac{\phi}{\Lambda_{\gamma}}\right), \hspace{1 em} m_e' = m_e\left(1+\frac{\phi}{\Lambda_e}\right).
\end{equation}

The size and refractive properties of solids depend on the fine structure constant and the electron mass.
We consider the effect of scalar field DM on the optical parameter

\begin{equation}
    \beta=\frac{2\pi d\Delta n}{\lambda},
\end{equation}

which is the difference in the accumulated phase between orthogonal polarizations in radians, where $d$ is the extent of the solid along the optical path, $\Delta n = n_e-n_o$ is the intrinsic birefringence of the solid (i.e. the difference of the refractive indices for the two orthogonal polarizations) and $\lambda$ is the wavelength of light. Relative changes in $\beta$ are the sum of relative changes in $d$ and $\Delta n$

\begin{equation}
   \frac{\delta \beta}{\beta} = \frac{\delta d}{d} + \frac{\delta \Delta n}{\Delta n},
\end{equation}

to first order. The size of a solid is proportional to the atomic Bohr radius, i.e $d\propto a_B = 1/(m_e\alpha)$ \cite{pasteka_material_2019}, where $\alpha$ is the fine structure constant and $m_e$ is the electron mass and therefore \cite{grote_novel_2019}

\begin{equation}\label{eq:deltad}
    \frac{\delta d}{d}=-\left(\frac{\delta \alpha}{\alpha}+\frac{\delta m_e}{m_e}\right)\left(1-\frac{\omega^2}{\omega_0^2}\right)^{-1}
\end{equation}

where we assume a high mechanical quality factor of the fundamental longitudinal vibrational resonance of the solid $\omega_0$ (i.e., weak damping)~\cite{aiello_constraints_2021}. The refractive index depends on the electronic resonances of the solid. If we only consider frequencies far away from the nearest electronic resonance $\omega_\phi\ll\omega_e$, the index of refraction is approximately inversely proportional to the electronic resonance of the solid, i.e. $1/n \propto \omega_e \propto m_e\alpha^2$, and so \cite{grote_novel_2019}

\begin{equation}\label{eq:deltan}
    \frac{\Delta n}{n}\approx C \left(2\frac{\delta \alpha}{\alpha}+\frac{\delta m_e}{m_e}\right),
\end{equation}

 where $C= \omega/n \cdot \partial n/\partial \omega$ takes into account the chromatic dispersion (see \cite{grote_novel_2019} for a more careful consideration of the approximations used in deriving Eqs.~\ref{eq:deltad},~\ref{eq:deltan}).

In summary, we thus expect that in the presence of a hypothetical scalar field there will be oscillatory changes of the parameter $\beta$:

\begin{eqnarray}
\nonumber \frac{\delta\beta}{\beta} &=& \phi_0\cos(\omega_{\phi}t-\vec{k}_{\phi}\cdot\vec{r})\times\\
&& \left[ \frac{\left(\frac{1}{\Lambda_\gamma} + \frac{1}{\Lambda_{e}}\right)}{ \left|1-\left(\frac{\omega_\phi}{\omega_0}\right)^2\right|}+ C\left(\frac{1}{\Lambda_{e}} + \frac{2}{\Lambda_\gamma} \right) \right].
\label{eq:signal}
\end{eqnarray}

\subsection{Polarization Rotation due to Axion DM}
Light pseudoscalar fields, including the axion and other axion-like particles, manifest themselves as an oscillating classical field, analogous to the scalar case \cite{antypas_new_2022, marsh_axion_2016, sakharov1996large, berezhiani1991cosmology, berezhiani1992primordial}:

\begin{equation}\label{eq:DMfield_ax}
    a(t,\vec{r})=a_0\,\cos(\omega_{a}t-\vec{k}_{\phi}\cdot\vec{r}),
\end{equation}

where $\omega_{a}=(m_{a}c^2)/\hbar$ is the angular Compton frequency for an axion mass $m_a$, the amplitude $a_0=(\hbar\sqrt{2\rho_{\mathrm{local}}})/(m_{a}c)$, and the other variables are the same as for Eq.~\ref{eq:DMfield}. We again assume the undiscovered field accounts for all of the local dark matter density.

We consider the coupling of the axion to the photon field parameterised by $g_{a\gamma}$ \cite{derocco_axion_2018}

\begin{equation}\label{eq:lagrangian_ax}
    \mathcal{L}_{\mathrm{int}}=\frac{a}{g_{a\gamma}}\frac{F_{\mu\nu}\tilde{F}^{\mu\nu}}{4}
\end{equation}

where $\tilde{F}^{\mu\nu}=\epsilon^{\mu\nu\rho\sigma}F_{\rho\sigma}$. Due to this coupling of photons with the axion field, there would be a difference in the phase velocity of right- and left-handed circularly polarized light \cite{nagano_axion_2019}: 

\begin{equation}
    v_{\circlearrowright,\circlearrowleft} \approx 1 \pm \frac{g_{a\gamma}\dot{a}}{2 k}
\end{equation}

Therefore, the right- and left-hand circular polarization components of light accumulate a relative phase difference, which produces a rotation of the plane of polarization of linearly polarized light by an angle \cite{martynov_quantum-enhanced_2020} 

\begin{equation}
    \rho(t,\tau) = \frac{g_{a\gamma}}{2}\left[a(t) - a(t-\tau)\right],
\end{equation}

for light propagating between times $t-\tau$ and $t$. For propagation times $\tau\ll1/\omega_a$, we have

\begin{equation}
    \rho(\omega_a,\tau,t) \approx g_{a\gamma} a_0 \omega_a \tau \sin(\omega_a t),
    \label{eq:axion_osc}
\end{equation}
to first order.

\section{Method}
\label{sec:method}

\subsection{Polarimetry setup for scalar field DM}

\begin{figure}[htb]
\begin{center}
\includegraphics[width=7.5cm]{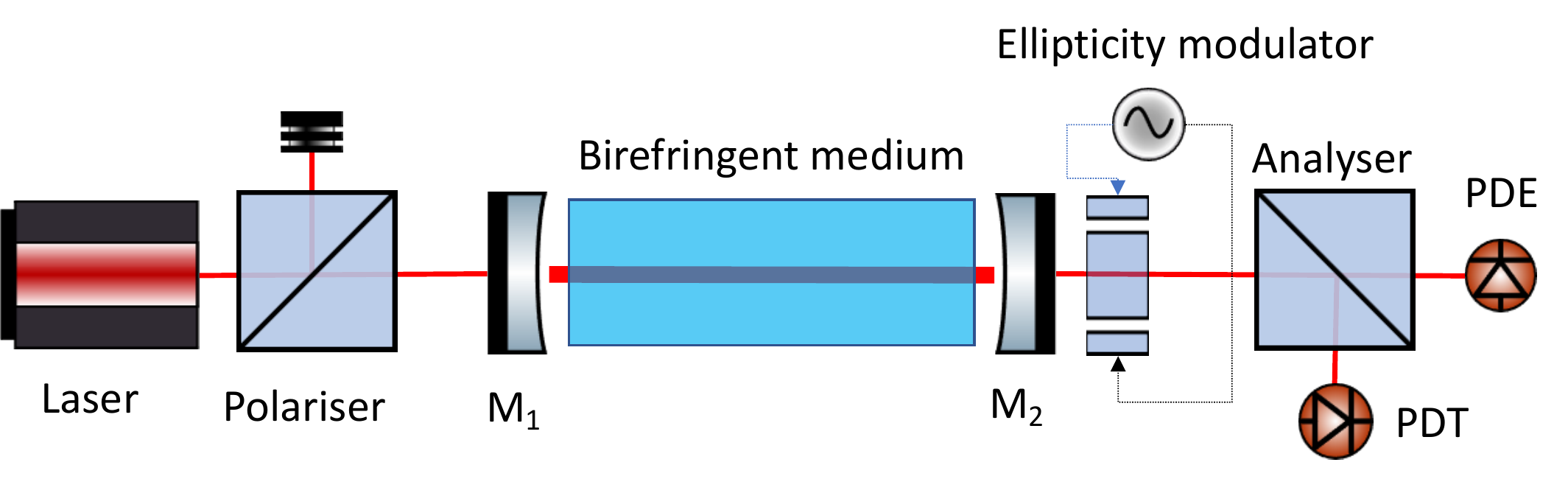}
\end{center}
\caption{The proposed polarimetry scheme for measuring the oscillation of scalar field dark matter using a birefringent medium in a Fabry-Pérot cavity. M$_1$/M$_2$ are mirrors that delimit the Fabry-Pérot cavity, and PDT/PDE are the photodiodes in the transmission and in the extinction port of the analyzer, respectively.}
\label{scheme-newSDM}
\end{figure}

In this section, we present an analytical calculation that shows how thickness variations of a birefringent optical element in a polarimeter, induced by scalar field DM, would produce a measurable phase difference between orthogonal polarization components of laser light (see Fig.\ref{scheme-newSDM}). 

To represent the optical field including its polarization, the Jones matrix formalism is used. After passing through a polarizer, the input laser light will be in the following state 

\begin{equation}
   \bm{\mathcal{E}}=E_0\,e^{- i (k z - \omega  t)}\cdot\left(
\begin{array}{c}
 1 \\
 0 \\
\end{array}
\right) 
\end{equation}

where $E_0$ is the electric field's initial magnitude. The Jones matrix of a birefringent solid with an optic axis oriented at an angle $\phi$ relative to the input polarization is:

\[
\mathbf{B}(\beta, \phi)= \mathbf{O}(\phi) \cdot \left(
\begin{array}{cc}
 e^{\frac{1}{2} i \beta} & 0 \\
 0 & e^{-\frac{1}{2} i \beta} \\
\end{array}
\right)\cdot  \mathbf{O}(-\phi)
\]

where $\mathbf{O}(\phi)$ is the rotation matrix:

\[
\mathbf{O}(\phi)=\left(
\begin{array}{cc}
 \cos (\phi ) & -\sin (\phi ) \\
 \sin (\phi ) & \cos (\phi ) \\
\end{array}
\right).
\]

A photoelastic modulator (PEM) and an analyzer are represented by the Jones matrices $\mathbf{H}$ and $\mathbf{A}$, respectively, 

\[
 \mathbf{H}=\left(
\begin{array}{cc}
 1 & i \eta  \\
 i \eta  & 1 \\
\end{array}
\right),\; \mathbf{A}=\left(
\begin{array}{cc}
 1 & 0 \\
 0 & 1 \\
\end{array}
\right).
\]

If the birefringent solid has an optic axis oriented at an angle of $\phi=\pi/4$, and the total phase-induced difference is $\beta=n\pi + \delta\beta, \; n\in\mathbb{Z}$, where $\delta\beta\ll1$ represents the phase modulation due to solid thickness and refractive index variation, the electric field after multiple reflections in the Fabry-Pérot cavity is as follows:

\begin{eqnarray*}
\mathbf{E}_{\rm out}&=&\left(\begin{array}{c}E_{{\rm out},0}\\E_{{\rm out},\perp}\end{array}\right)\\
&=&  Te^{\frac{i \varphi}{2}}\,\sum_{n=0}^\infty\left[Re^{i \varphi}\,\mathbf{B}^2(\beta, \frac{\pi}{4})\right]^n \cdot \,\mathbf{B}(\beta, \frac{\pi}{4})\cdot\bm{\mathcal{E}}\\
&=& Te^{\frac{i \varphi}{2}}\,\left[\mathbf{I}-Re^{i \varphi}\,\mathbf{B}^2(\beta, \frac{\pi}{4})\right]^{-1}\cdot\,\mathbf{B}(\beta, \frac{\pi}{4})\cdot\bm{\mathcal{E}}
\end{eqnarray*}

where $\varphi$ is the round-trip phase shift of light propagating between the two-cavity mirrors, $R$ is the reflectance of the mirrors, $T$ is the transmittance of the mirrors, and $\mathbf{I}$ represents the identity matrix. 

The heterodyne method involves placing a PEM next to an analyzer oriented at $45^\circ$ with respect to the input polarization (see Fig.~\ref{scheme-newSDM}). The electric field at the extinguished port of the analyzer is:

\begin{equation}
  \mathbf{ E}_\mathrm{ext}=\mathbf{ A}\cdot\mathbf{H}\cdot\mathbf{E}_{\rm out}.
\end{equation}

The following equation is used to compute the extinguished intensity: $I_{\rm ext} = |\mathbf{ E}_\mathrm{ext}|^2$. The expression for relative intensity, taking into consideration the extinction ratio $\sigma$ of the two polarizers, is:

\begin{eqnarray}
&\frac{I^{\rm ext}}{I_0} &=\displaystyle\frac{T^2}{1-2 R \cos\varphi+R^2}\\ 
\nonumber &\times&\left(\sigma^2+ \eta^2+\eta~\delta\beta\frac{\left(1-R^2\right)}{1-2 R \cos\varphi+R^2}+ \mathcal{O}\left[(\delta\beta)^2 \right]\right)
\label{eq:intSDM}
\end{eqnarray}

where $\delta\beta$ represents the phase difference due to the birefringent solid thickness and refractive index variation, and ${\eta(t) = \eta_0 \cos (2\pi\nu_{\rm PEM}t)}$ and $\sigma$ represent the extinction of the crossed polarizers. The term linear in $\eta(t)$ is the heterodyne signal, which is used to detect the effect of interest. Furthermore, we must consider that the Fabry-Pérot cavity operates as a low-pass first-order filter for a time-dependent signal such as the one induced by scalar field DM. Specifically, the transfer function of a Fabry-Pérot cavity for a time-dependent ellipticity signal is \cite{ejlli_polarisation_2018}

\begin{eqnarray}
\nonumber
h_{\rm T}(\nu)&=&\frac{T}{\sqrt{1+R^2-2R\cos2\pi\nu\tau}}\\
\phi_{\rm T}(\nu)&=&\arctan\left[\frac{R\sin2\pi\nu\tau}{1-R\cos2\pi\nu\tau}\right],
\label{eq:First}
\end{eqnarray}

where $\nu$ is the frequency of the signal, ${\tau=2/c \int^{L}_{0} ndL}$ is one round trip time in the cavity, and $n$, $L$, and $c$ are the refractive index, the length of the cavity, and the speed of light, respectively.

The input laser light should be kept resonant in the Fabry-Pérot cavity, such that $\varphi=0$, using e.g. the Pound-Drever-Hall method. The signal from the photodiode at the extinguished port can be demodulated at the frequency $\nu_{\rm PEM}$, and the magnitude of the phase shift $|\delta\beta|$ can then be inferred:

\begin{equation}
   |\delta\beta(\nu)| =  \frac{\mathcal{I}^{\rm ext}_{\nu_{\rm PEM}}(\nu)}{N I_0 \eta_0 h_{\rm T}(\nu)},
    \label{eq:signal_dem_pemSDM}
\end{equation}

where $N=2/(1-R)$ is the cavity buildup and $\mathcal{I}^{\rm ext}_{\nu_{\rm PEM}}(t)$ is the demodulated signal of the extinguished intensity at frequency $\nu_{\rm PEM}$.

\subsection{Polarimetry for axion DM}
We now consider the case of pseudoscalar axion-like DM and show how such a field may produce an observable signal in a polarimeter. 

\begin{figure}[htb]
\begin{center}
\includegraphics[width=7.5cm]{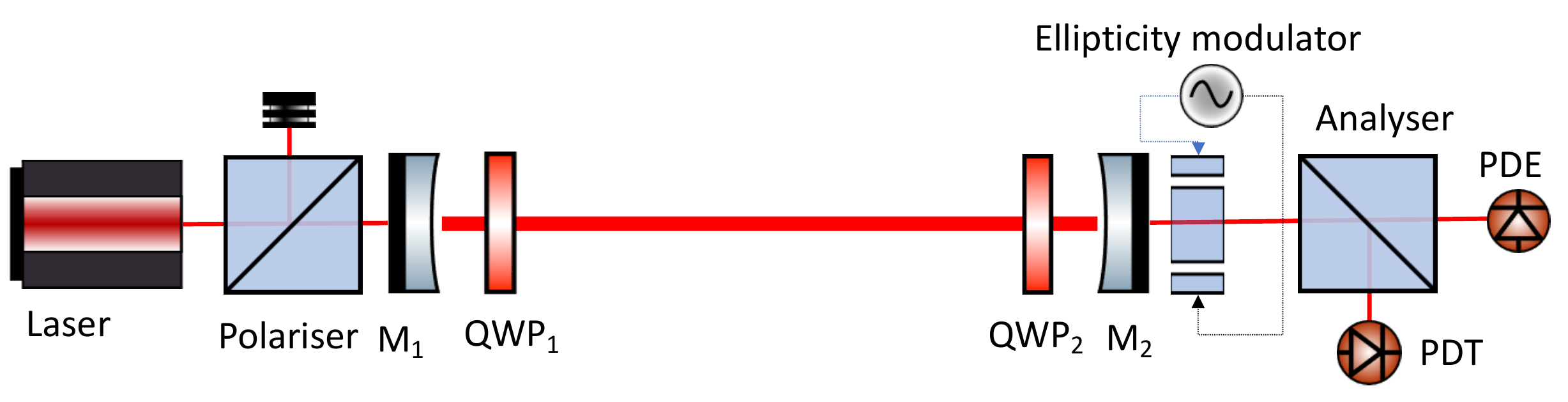}
\end{center}
\caption{The proposed polarimetry scheme for sensing the coupling of the axion field to photons by using a Fabry-Pérot cavity and two quarter-wave plates (QWP$_1$ and QWP$_2$). PDT/PDE are the photodiodes in the transmission and in the extinction port of the analyzer, respectively, and M$_1$/M$_2$ are mirrors that delimit the Fabry-Pérot cavity.}
\label{scheme-newADM}
\end{figure}

The Jones matrix (on a linear basis) for the propagation of light that parameterises the rotation of its plane of polarization in the presence of an axion field is given by

\begin{equation}
\mathbf{A_\gamma}(t,\tau)\approx\left(
\begin{array}{cc}
 1 & -\rho(t,\tau)/2  \\
 \rho(t,\tau)/2  & 1\\
\end{array}
 \right),
\end{equation}

for angles $\rho \ll 1$. We first consider a setup where polarized laser light is injected into an empty Fabry-Pérot cavity, without any optical elements inside. The rotation of the plane of polarization of the light inside the cavity due to an axion field is

\begin{eqnarray*}
\mathbf{E}_{\rm out}&=& \sum_{n=0}^\infty\left[e^{i \varphi}\,(\mathbf{A_\gamma}(\rho)\cdot\mathbf{M})^2\right]^n \cdot Te^{i \varphi/2}\,\mathbf{A_\gamma}(\rho)\cdot\bm{\mathcal{E}}\\
&=& \frac{Te^{i \varphi/2}}{(1-Re^{i \varphi})}\,\mathbf{A_\gamma}(\rho)\cdot \bm{\mathcal{E}}
\end{eqnarray*}

to first order, where $\mathbf{M}_{\rm } = \left(\begin{array}{cc} -r & 0  \\ 0  & r\\\end{array} \right)$ is the Jones matrix representing the reflection of a mirror for normal incidence and $\varphi$ is the round trip phase shift. 

We note that over a round trip the element $(\mathbf{A_\gamma}(t,\tau)\cdot\mathbf{M})^2= R\,\mathbf{I}$, where $R=r^2$ is the reflectance of the mirror and $\mathbf{I}$ is the identity matrix. Therefore, when the light is resonant with the cavity, the axion-induced polarization polarization rotation, also known as dichroism, cancels out over round trips.

To eliminate this cancellation effect caused by the round trip, two quarter-wave plates (QWPs) can be placed next to the mirrors at either end of the cavity, as shown in Fig.~\ref{scheme-newADM}. This is similar to the approaches in \cite{derocco_axion_2018,martynov_quantum-enhanced_2020}.  The fast axes of the QWPs must be aligned at an angle of $\phi = (0,\, \pi/2)$ with respect to the input polarization plane. In this configuration, the polarization rotation due to the axion field will accumulate over multiple round trips. A quarter-wave plate with its fast axis aligned with the incident polarization is represented by the matrix $\mathbf{Q}=\mathbf{B}(\pi/2,0)$. The quarter-wave plate would cause the rotation effect on the polarization of light passing through it to become elliptical:

\begin{equation}
\mathbf{Q}\cdot\mathbf{A_\gamma}\cdot\mathbf{Q}\cdot\bm{\mathcal{E}} = \Tilde{E_0}
\left(
\begin{array}{c}
 1 \\
 -\frac{i \rho}{2} \\
\end{array}
\right),
\end{equation}

where  $\Tilde{E_0}= E_0 e^{-i {\zeta}}$ is the amplitude of the electric field with an overall phase $\zeta$.

The electric field at the exit of the Fabry-Pérot cavity including the QWPs is:

\begin{eqnarray*}
\mathbf{E}_{\rm out}
&=&  Te^{\frac{i\varphi}{2}}\sum_{n=0}^\infty\left[e^{i \varphi}\,\left(\mathbf{Q}\cdot\mathbf{A_\gamma}(\rho)\cdot\mathbf{Q}\cdot\mathbf{M}\right)^2\right]^n\\ \nonumber&\cdot&\,\mathbf{Q}\cdot\mathbf{A_\gamma}(\rho)\cdot\mathbf{Q}\cdot \bm{\mathcal{E}}\\
&=& Te^{\frac{i \varphi}{2}}\left[\mathbf{I}-e^{i \varphi}\,\left(\mathbf{Q}\cdot\mathbf{A_\gamma}(\rho)\cdot\mathbf{Q}\cdot\mathbf{M}\right)^2\right]^{-1}\\
\nonumber&\cdot&\mathbf{Q}\cdot\mathbf{A_\gamma}(\rho)\cdot\mathbf{Q}\cdot \bm{\mathcal{E}}
\end{eqnarray*}

Using the heterodyne technique and taking into account the extinction ratio of the two polarizers, the intensity of the field at the extinguished port relative to the input intensity is

\begin{equation}
\frac{I^{\rm ext}}{I_0}(\rho) \approx \sigma ^2+\eta ^2-2N\eta\frac{\rho}{2}+O\left[\rho^2 \right].
\label{eq:intADM}
\end{equation}

When the frequency response (see Eq.~\ref{eq:First}) is taken into consideration, the phase shift due to the axion that can be measured is

\begin{equation}
   |\rho(\nu)| =  \frac{\mathcal{I}^{\rm ext}_{\nu_{\rm PEM}}(\nu)}{N h_{\rm T}(\nu) I_0 \eta_0}.
    \label{eq:signal_dem_pemADM}
\end{equation}

This expression shows that with the addition of the two quarter-wave plates, the phase shift between orthogonal polarizations induced by an axion field builds over multiple cavity trips $N$. The accumulated signal takes the form of elliptical polarization of the output light. 

Eqs.~\ref{eq:signal_dem_pemSDM} and \ref{eq:signal_dem_pemADM} show that a polarimeter can be used for a direct search for both scalar and pseudo-scalar dark matter, by configuring the device either as in Fig.~\ref{scheme-newSDM} or as in Fig.~\ref{scheme-newADM}.

The approach presented here shares similarities with the method described in Nagano et al.  \cite{nagano_axion_2019} for detecting axion dark matter using interferometric gravitational wave detectors. However, our method differs in the use of quarter-wave plates to accumulate the rotation induced by axion dark matter over multiple round trips and subsequently convert it into ellipticity for detection through polarimetry \cite{derocco_axion_2018}.

\section{Noise budget}
\label{sec:noise}

We carried out a preliminary study of noises that can limit the performance of the proposed polarimetry experiment, such as shot noise, Johnson noise, seismic noise, and limitations due to optical aberrations. The intrinsic noises that will limit the sensitivity can be calculated from equations~(\ref{eq:signal_dem_pemSDM}, \ref{eq:signal_dem_pemADM}) as follows:

\begin{equation}
S_{\rm P}=\frac{S_{\mathcal{I^{\rm ext}_{\nu_{\rm PEM}}}}}{I_0\eta_0}.
\end{equation}
There are different effects that generate noises that add in quadrature to the $S_{\mathcal{I^{\rm ext}_{\nu_{\rm PEM}}}}$, which can all be expressed in terms of the extinguished intensity $ I_{\rm ext} $.

\subsection{Shot noise}

Let us begin by analyzing the spectral density of the intrinsic RMS noise that is caused by the photodiode's direct current $i_{\rm dc}$

\begin{equation}
i^{\rm (shot)}=\sqrt{2e\,i_{\rm dc}},
\end{equation}

measured in ampere/$\sqrt{\rm hertz}$. Note that $i^{\rm (shot)}$ is independent of frequency and the direct current in the photodiode is $i_{\rm DC} = qI_0\eta_0^2/2$, where $q$ is the efficiency of the detector PDE in units of A/W. The well-known extinction ratio of two crossed polarizers introduces an additional component in the detected DC power, which is expressed as $I_0\sigma^2$. The extinction ratio of the polarizers is taken into account, which can be as low as $\sigma^2\approx 10^{-7}$. This leads to an expression for the shot-noise spectral densities of the light power $I^{\rm (shot)}$ and in the ellipticity $S_\beta^{\rm (shot)}$:

\begin{equation}
I^{\rm (shot)}=\frac{i^{\rm (shot)}}{q} = \sqrt{\frac{2e\,I_0}{q}\left(\sigma^2+\frac{\eta_0^2}{2}\right)},
\end{equation}

and

\begin{equation}
 S^{\rm (shot)}_{\beta}=\sqrt{\frac{2e}{qI_0}\left(\frac{\sigma^2+\eta_0^2/2}{\eta_0^2}\right)}.
\end{equation}

\subsection{Electronic noise and RIN}

Other noises that affect the spectral densities of ellipticity noise include Johnson noise of the photodiode transimpedance $G$,

\begin{equation}
I^{\rm (J)}=\sqrt{\frac{4k_BT}{q^2G}},\qquad S_{\rm P}^{\rm (J)}=\sqrt{\frac{4k_BT}{G}}\frac{1}{qI_0\eta_0},
\end{equation} 

the photodiode dark current

\begin{equation}
I^{\rm (dark)}=\frac{i_{\rm dark}}{q},\qquad{\rm with}\qquad S_{\rm P}^{\rm (dark)}=\frac{i_{\rm dark}}{qI_0\eta_0},
\end{equation}

and the frequency-dependent relative intensity noise $N^{\rm (RIN)}_{\nu}$ of the light emerging from the cavity is estimated as:

\begin{equation}
I^{\rm (RIN)}_{\nu}=I_0\,N^{\rm (RIN)}_{\nu},
\end{equation}

giving

\begin{equation}
S^{\rm (RIN)}_{\rm P}=N^{\rm (RIN)}_{\nu_m}\,\frac{\sqrt{(\sigma^2+\eta_0^2/2)^2+(\eta_0^2/2)^2}}{\eta_0}.
\label{eq:SRIN}
\end{equation}

\subsection{Seismic noise}

\begin{figure}[bht]
    \centering
    \includegraphics[width=7.5cm]{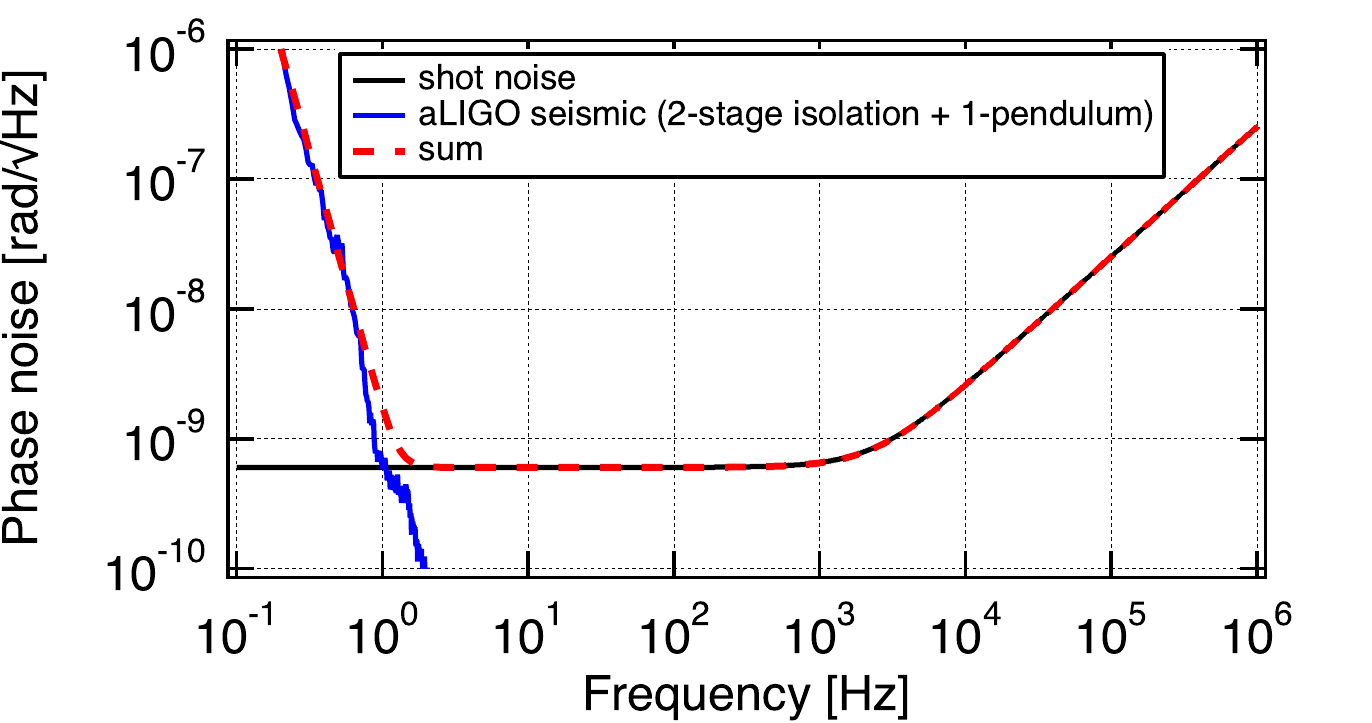}
    \caption{A projection of two forms of sensitivity phase noise (SNR=1) expected in the proposed polarimetry experiment are shown in this figure: shot and seismic sensitivity phase noises. For calculating the curves in this plot, we used Eq.~\ref{GradientDeltanD1}, and the parameters on Tab.~\ref{tab:1}. The seismic phase noise amplitude plotted in blue corresponds to the noise as mitigated through a two-stage seismic isolation platform and a single silica pendulum suspension, which is part of the seismic isolation and suspension system used at the advanced Laser Interferometer Gravitational-Wave Observatory (aLIGO) \cite{Schwartz_2020, LIGOreviewpaper}.}
    \label{fig:noise}
\end{figure}

Seismic noise could be significant in the proposed polarimetry scheme for detecting scalar field DM (Fig.~\ref{scheme-newSDM}) if the back and front surfaces of the birefringent crystal are not parallel and the point of incidence of the laser beam on the solid varies. In this situation, the laser light will scan various solid thicknesses, thereby generating unwanted polarization phase noise. If the optical components of the polarimeter are installed on a perfectly rigid platform, the seismic disturbance should have no effect on their relative position. Nevertheless, the coherence length of seismic noise above 1.5~Hz reduces to $1-16$~m \cite{satari_2022}, which would mean the cavity mirrors do experience a differential displacement.

To estimate the coupling of seismic noise into the polarimeter, we consider our birefringent solid to have uniform birefringence and a small wedge, $\theta$. We assume the solid's transverse position relative to the incidence of the laser beam is randomly modulated by the seismic noise $\delta r$. Further, we assume that the seismic noise is broadband and has the same magnitude in all directions. The coupling of the seismic noise to birefringence noise in the polarimeter can then be calculated using the wedge angle $\theta$ as follows:

\begin{equation}
S_{\rm P}^{\rm (seismic)}=2\pi \frac{N\, \delta r\, \theta}{\lambda}~\Delta n \gamma{\left(L,f\right)}
\label{GradientDeltanD1}
\end{equation}

Here, $\Delta n$ is the birefringence of the solid, $N$ is the cavity build-up, and $\gamma{(L,f)}$ represents the two-point correlation of the seismic noise at the laser and the solid separated by a distance $L$ as a function of frequency. This correlation is defined here such that if the seismic displacement of the birefringent solid and the laser beam are correlated, $\gamma=0$, which means the relative displacement of the laser and solid is zero. For $\gamma=1$ the relative displacement is completely uncorrelated and noise coupling is at its maximum.

We assume the operational parameters for the polarimeter have the values detailed in Table~\ref{tab:1}, which are typical operating characteristics for a polarimetry setup~\cite{ejlli_pvlas_2020}. We estimate a shot noise that is greater than both the electronic noise and the RIN, with a magnitude of $S_{\rm shot}^{\rm (tot)}\approx6\times10^{-10}$.

The total ellipticity noise is expected to be limited by the sum in quadrature of shot noise and seismic noise:

\begin{equation}
 S_{\rm P}^{\rm (tot)} \approx \sqrt{{S_{\rm P}^{\rm (shot)}}^2+{S_{\rm P}^{\rm (seismic)}}^2+ {S^{\rm (RIN)}_{\rm P}}^2 +{S^{\rm (dark)}_{\rm P}}^2}
\label{eq:tot_noise}   
\end{equation}

 where shot and seismic noise are expected to be the dominant noises in the total noise budget.

\begin{table}[htbp]
\centering
\begin{tabular}{|c|c|c|}
\hline
       Input power               & $I_0$          & $1$~W     \\ \hline
        PDE quantum efficiency   &$q $            & 0.7~A/W   \\ \hline
        PDE gain                 &$G$             &$10^6~{\rm\Omega}$ \\ \hline
        Extinction ratio         &$\sigma^2$       &$2\times10^{-7}$  \\ \hline
        Dark noise               &$i_{\rm dark}$  &25~fA$_{\rm rms}/\sqrt{\rm Hz}$ \\ \hline
        Modulation amplitude     & $\eta_0$           & $1.5\times10^{-3}$    \\ \hline
        Modulation frequency     & $\nu_{\rm PEM}$            & $50$~kHz    \\ \hline
        RIN  \cite{mephisto_lasers, ejlli_pvlas_2020}             &$N^{\rm (RIN)}_{\nu_{\rm PEM}}$   &$3\times10^{-7}/\sqrt{\rm Hz}$ \\ \hline
        Seismic noise coupling &$\gamma$                &0.1 \\ \hline
        Cavity build-up                &$N$             &20\,000 \\ \hline
        Solid/QWP wedge                   &$\theta$         & 1~$\mu$rad \\ \hline
        Yttrium Vanadate                   &$C$         & $12\times 10^{-3}$ \\ \hline
        Sapphire                  &$C$         & $6.6\times10^{-3}$ \\ \hline
    
\end{tabular}
\caption{Relevant characteristics of the proposed polarimetry setups}
\label{tab:1}
\end{table}

In Fig.~\ref{fig:noise} we show the phase noise sensitivities for a signal-to-noise ratio (SNR) of 1, limited by shot noise and seismic noise. The seismic phase noise amplitude plotted in blue corresponds to the noise as mitigated through a two-stage seismic isolation platform and a single silica pendulum suspension, which is part of the seismic isolation and suspension system used at the advanced Laser Interferometer Gravitational-Wave Observatory (aLIGO) \cite{Schwartz_2020, LIGOreviewpaper}. From this figure, we can clearly see that at sub-Hertz frequencies, seismic noise becomes the dominant factor in our proposed scheme.

\subsection{Optical aberrations}

In order to maintain the required sensitivity, the Fabry-Pérot cavity must be kept on resonance with the laser light, and the light at the extinguished port of the analyzer must be kept to a minimum. To satisfy both these conditions in the polarimetry configuration for scalar field DM (Fig.~\ref{scheme-newSDM}), it is necessary that the total relative phase induced by the birefringent solid is $\beta=n\pi, \; n\in\mathbb{Z}$. On the other hand, in the configuration for axion-like particles (Fig.~\ref{scheme-newADM}), the total relative phase retardation imparted by the QWPs must be exactly $\pi/2$. 

Given these constraints, control over thermal effects is required in both cases since these impact the optical path length inside birefringent solids. The presence of a nonzero thermal optical coefficient $dn/dT\neq 0$ causes changes in the refractive index of an optic due to the absorption of optical power in the substrate and in the coatings. This results in the formation of a thermal gradient inside the optic, creating an optical aberration called thermal lensing. In addition, due to the non-zero thermal expansion coefficient $\chi$, the surface of a heated optic expands along the optical axis and induces a further change in the optical path length known as thermo-elastic deformation.
The total variation of the optical path due to these effects is calculated as \cite{Rocchi, AielloPhD}

\begin{equation}\label{eq:thermal}
    \delta(n d)=\left( \frac{dn}{dT}\Delta T + \chi(1+\epsilon)(n-1)\Delta T \right) d
\end{equation}

where $\epsilon$ represents Poisson's ratio. However, the contribution from thermo-elastic deformation is usually small compared to that from thermal lensing \cite{Rocchi}. A crystal placed inside a Fabry-Pérot cavity with thermal lensing in its Anti-Reflection (AR) coatings and substrates presents a problem similar to those encountered in e.g. interferometric gravitational-wave detectors, and we use the analysis methods and solutions used in those instruments for this work. 

In order to estimate the thermal lensing and the related variation in the optical path length inside the crystal in the proposed polarimetry configurations, finite element analysis simulations based on the Hello-Vinet theory \cite{Hello:Vinet} have been carried out with FINESSE~3 \cite{Finesse3}. For these calculations, we assumed the input laser light to be purely in the fundamental Hermite-Gauss~00 mode with a beam radius of $w\sim 1$~cm. We performed simulations for two different birefringent materials; yttrium vanadate and sapphire, where the dimensions of the solid were set to 5~cm in thickness and 10~cm in width. 

\begin{figure}[htpb]
    \centering
    \includegraphics[width=7.5cm]{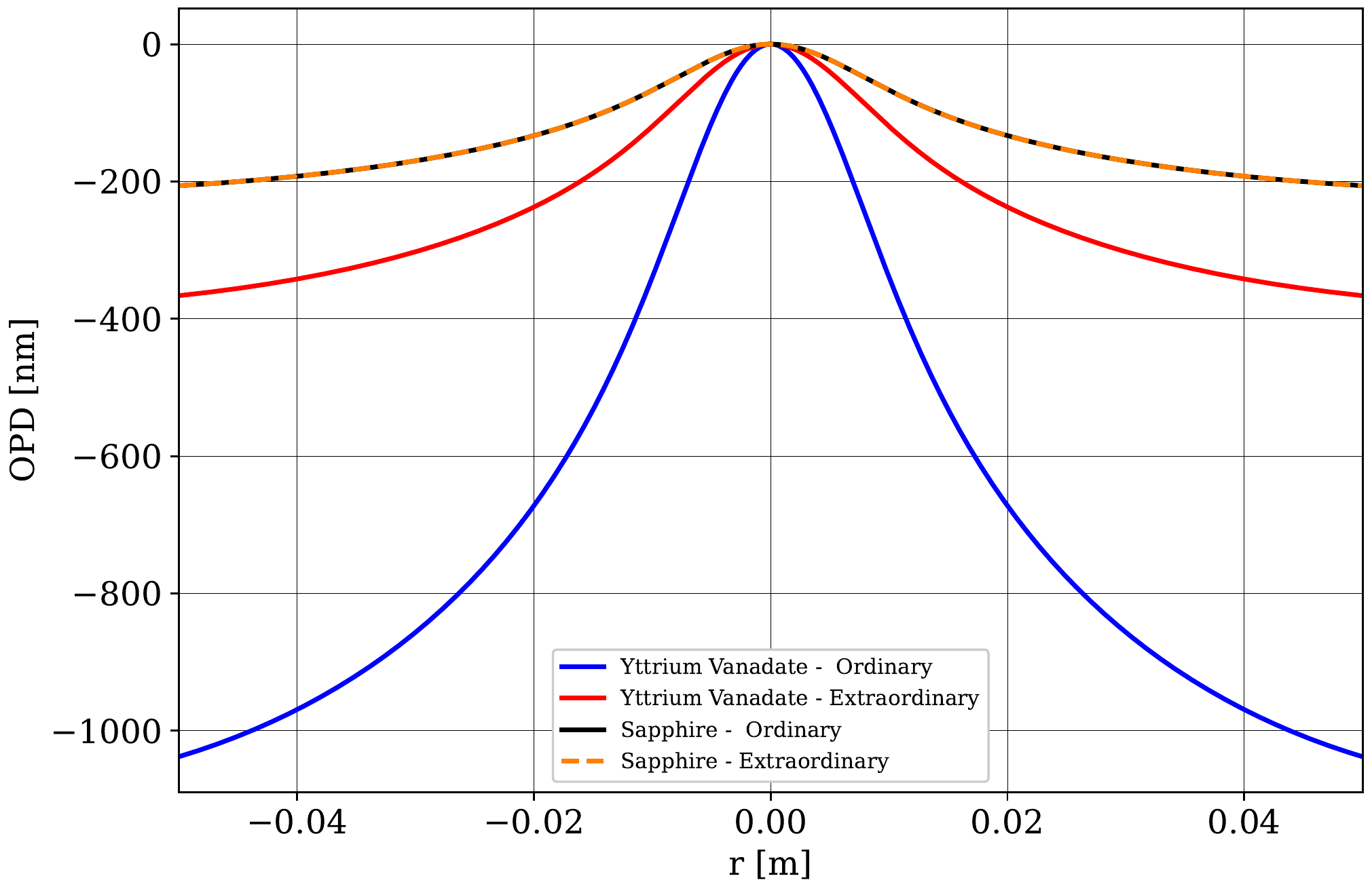}
    \caption{Simulated optical path difference as a function of the optic's radial coordinate generated by thermal lensing for Yttrium-Vanadate and Sapphire, assuming a bulk absorption of 100~pm/cm \cite{Marchio:2021xco}: $r=0$ represents the center of the solid where the incident beam enters.}
    \label{fig:OPD}
\end{figure}

In Fig.~\ref{fig:OPD}, we show the simulated optical path difference due to thermal effects (as given by Eq.~\ref{eq:thermal}) for both materials for the ordinary and extraordinary axes. We considered an input power of 1~W and a cavity buildup of 4000. It can be seen from these results that the thermally produced optical path difference (OPD) difference between the center and edge of the beam will be less than 1~$\mu$m. An OPD less than or of the order of 1~$\mu$m can be compensated by a specialized thermal actuator, like a CO$_2$ laser projector or a ring heater \cite{Hello_2001,Aiello_2019,AielloPhD}.

Because the thermal effects can affect the optical path length along the ordinary and extraordinary optic axis differently, the solid's birefringence parameter will change: $\beta = n\pi+ \delta\beta_{\rm T}$, where $\delta\beta_{\rm T}$ can range from 0 to $\pi/2$. The correction that needs to be effected to compensate, ($\delta\beta_{\rm T}$), will depend on the differential OPD between the extraordinary and ordinary axes. However, as we only need to correct the relative OPD to the nearest integer multiple of the wavelength, the needed correction is by definition less than 1~$\mu$m. 

For the configuration for axion-like particles, we estimate that the QWPs have a bulk absorption that is more than two orders of magnitude lower than that of the thick birefringent solid, which could easily be compensated for. 

\subsection{Calibration}

The phase and amplitude response of a polarimeter can be accurately calibrated using the Cotton-Mouton effect, that is the magnetic birefringence of gases. The birefringence generated in a gas at pressure $p$ by a magnetic field $B_{\rm ext}$ is given by the expression:

\begin{equation}
\Delta n = n_\parallel - n_\perp = \Delta n_u\, B_{\rm ext}^2\,p,
\end{equation}

where $\Delta n_u$ is the unitary birefringence at 1 atmosphere of pressure and 1~T of the magnetic field. Typical values of $\Delta n_u$ range from about $2.2\times10^{-16}$~T$^{-2}$atm$^{-1}$ for He to about  $-2.3\times10^{-12}$~T$^{-2}$atm$^{-1}$ for O$_2$ and to $\approx 10^{-11}$~T$^{-2}$atm$^{-1}$ for a few other simple molecules \cite{ejlli_pvlas_2020, della_valle_measurement_2014}. By filling the vacuum system housing the experiment with, e.g., O$_2$, and applying a magnetic field over some length of the cavity, accurate calibration can be performed. 

\section{Prospects}
\label{sec:Prospects}

\subsection{Scalar field DM}

In the proposed polarimeter setup for detecting scalar field DM, the sensitivity is defined by $\delta \beta / \beta$. From the noise budget, we see that the condition where the shot noise readout phase is the dominant noise, i.e. $\delta \beta= S_{\rm P}^{\rm (tot)}\approx 6\times 10^{-10}~/\sqrt{\rm Hz}$ can be achieved. The sensitivity to scalar field DM can be improved by increasing $\beta={ 2\pi d_{\rm eff}\Delta n/\lambda}$. The effective length of the solid $d_{\rm eff}$ in the cavity depends on the product of the solid thickness and the average number of round trips in the cavity. We evaluate the dependence of $\beta$ on the experimental conditions:

\begin{eqnarray}
\beta(\nu)&=& 2\pi \frac{ d (\nu)\Delta n}{\lambda}Nh_{\rm T}\\ 
\nonumber&=& \frac{2\pi d \Delta n}{\lambda} \frac{2}{ (T+P)\sqrt{1+\frac{4}{(T+P)^2} \sin^2{\pi \nu \tau}}}\\
\nonumber&\approx&\frac{2\pi d \Delta n}{\lambda} \frac{2}{\sqrt{P^2+4 \sin^2{\pi \nu \tau}}} \quad \text{for}\quad T\ll P.
\end{eqnarray}

In this approximation, we have assumed that absorption is the main loss, limiting the number of round trips, i.e. $P\gg T$, which would be the case for this experiment. Losses that may occur in the cavity are due to imperfections of the AR coating and the solid's bulk absorption loss $P=P_{\rm AR}+\mu d$, where $\mu$ is the linear absorption coefficient of the bulk. The sensitivity to scalar-field DM is 

\begin{equation}
    \frac{\delta \beta}{\beta} =\frac{S_{\rm P}^{\rm (tot)} \lambda \sqrt{P^2+4 \sin^2{\pi \nu \tau}}}{2\pi d \Delta n}.
    \label{eq:delta_beta_beta}
\end{equation}

According to the equation above, the sensitivity depends on the performance of the polarimeter, parameterized by $S_{\rm P}^{\rm (tot)}$, and the properties of the solid. Specifically, the sensitivity scales linearly with the differential optical path $d \Delta n$ and the total absorption caused by the solid $P$. As a consequence of this, if coating loss is the dominant effect, the sensitivity will be proportional to the birefringence of the material as well as its thickness. If, on the other hand, the bulk absorption contributes the majority of the total loss, the sensitivity will not improve with increased thickness. Therefore, the cross-over point between sensitivity gain due to increased thickness and loss due to bulk absorption will be the most important consideration when choosing the optimal solid thickness.

\begin{figure}[htpb]
    \centering
    \includegraphics[width=7.5cm]{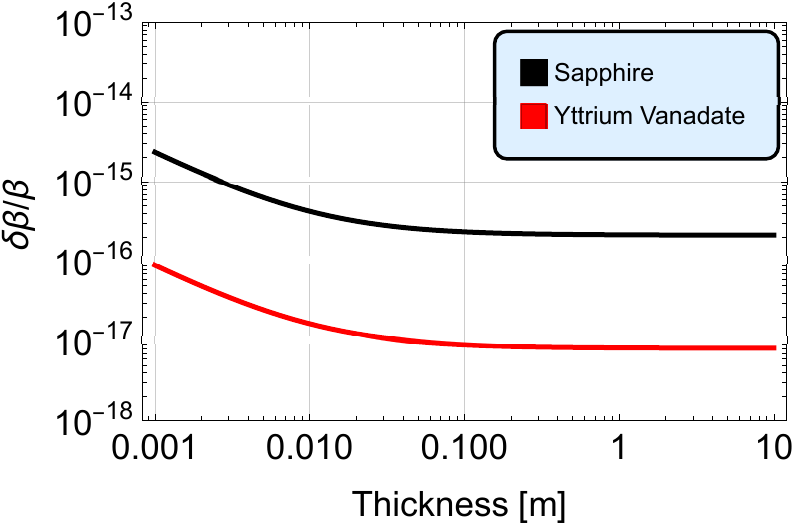}\\\includegraphics[width=7.5 cm]{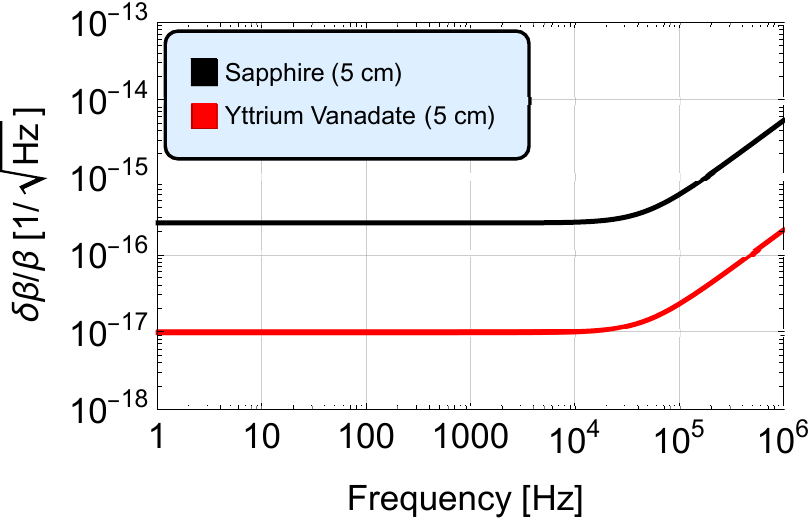}
    \caption{Upper panel: sensitivity of the polarimeter Eq.~\ref{eq:delta_beta_beta} as a function of the crystal thickness for Sapphire ($\Delta n=0.008$) and pure Yttrium Vanadate ($\Delta n =0.208$) with absorption of $P= 100$~ppm/cm. Bottom panel, sensitivity Eq.~\ref{eq:delta_beta_beta} of Yttrium Vanadate (YVO4) crystals and Sapphire as a function of frequency for a $d=5$ cm thick crystal.}
    \label{fig:sensitivity}
\end{figure}

Fig.~\ref{fig:sensitivity} shows the expected $\delta \beta/\beta$ sensitivity as a function of crystal thickness and frequency for two types of birefringent crystals: sapphire ($\Delta n=0.008$) and pure Yttrium Vanadate ($\Delta n =0.208$) crystals. These calculations were made using a linear bulk absorption coefficient of 100~ppm/cm \cite{Marchio:2021xco}, and a AR coating of 25 ppm per incidence. We have used a Fabry-Pérot cavity length of 30~cm and a thickness of 5~cm for the birefringent solid. Because bulk absorption increases linearly with crystal thickness, a crystal thicker than 5~cm would not yield a meaningful improvement.

\begin{figure}
    \centering
    \includegraphics[width=7.5cm]{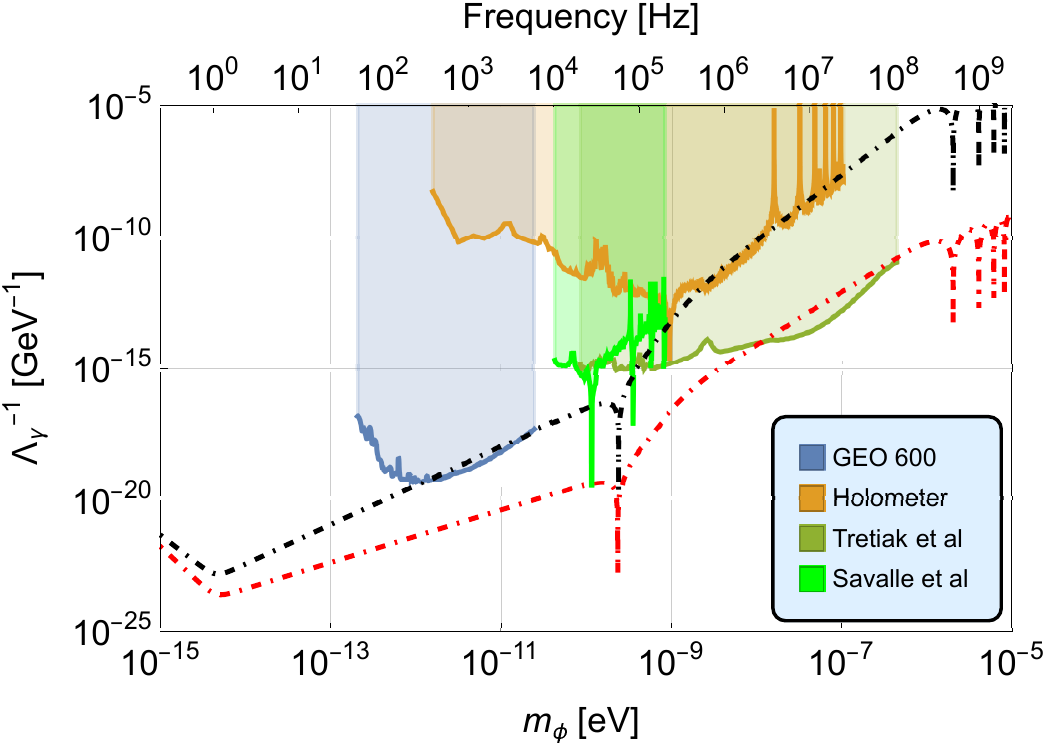}\\\includegraphics[width=7.5 cm]{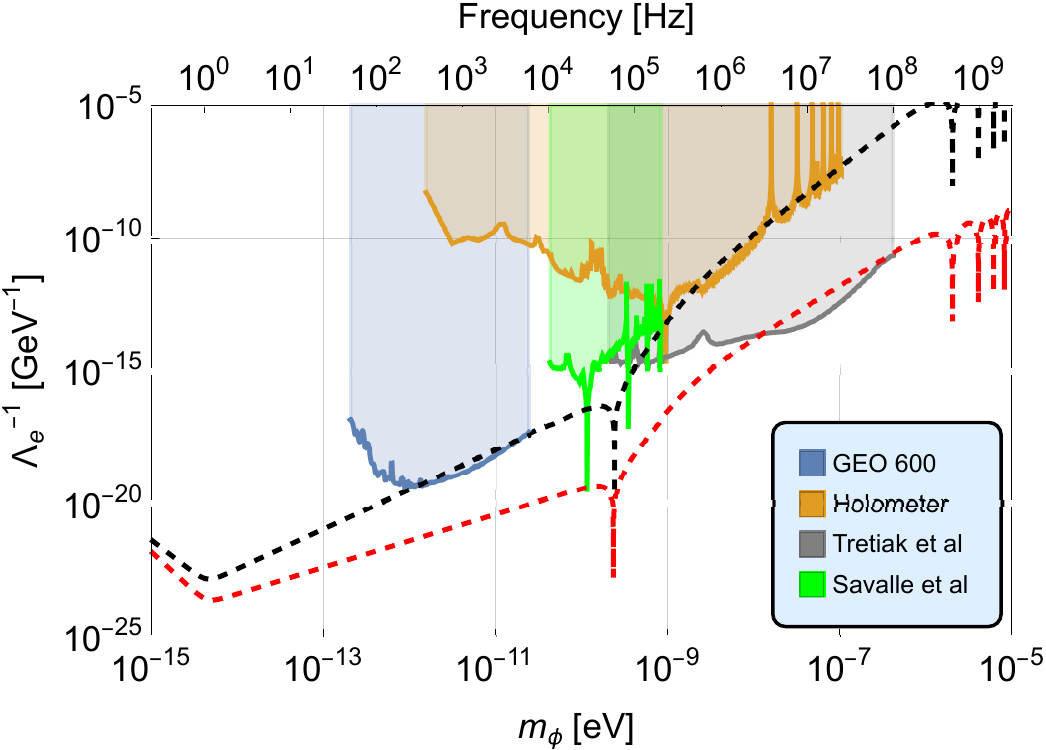}
    \caption{Prospects for sensitivity to scalar field dark matter using polarimetry with a 30-centimeter-long Fabry-Pérot cavity and a 5-centimeter-thick Yttrium Vanadate birefringent crystal are shown; the lines give the upper sensitivity in terms of the electron coupling constant (upper panel) and photon coupling constant (bottom panel), as a function of scalar field mass. The black dashed/dotted-dashed lines represent the integrated sensitivity over the coherence time of dark matter, while the red dashed/dotted-dashed lines represent one year of integration using twin polarimetry and cross-correlation. Existing constraints from other interferometry experiments are \cite{vermeulen2021direct,aiello_constraints_2021,savalle_searching_2021,tretiak_improved_2022} are also shown.}
    \label{fig:sensitivity_dark_matter}
\end{figure}

Prospects for scalar field DM are shown in Fig.~\ref{fig:sensitivity_dark_matter}, with electromagnetic coupling (bottom) and electron coupling (top) for a 5-cm Yttrium Vanadate. The black dashed lines represent the improved sensitivity through integration for the coherence time of DM, while the red dashed lines represent the integration using twin polarimeters cross-correlated for a total of one year. Existing constraints from other interferometry experiments \cite{vermeulen2021direct,aiello_constraints_2021,savalle_searching_2021,tretiak_improved_2022} are shown for comparison.

\subsection{Axion DM}
\begin{figure}[ht!]
    \centering
    \includegraphics[width=7cm]{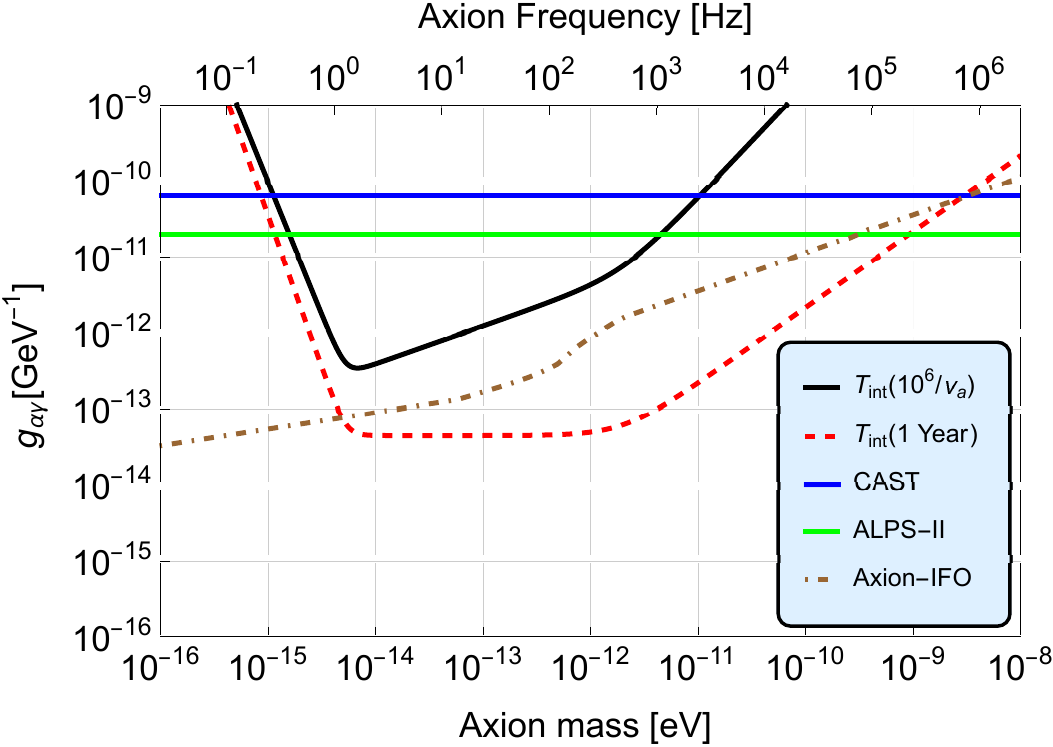}
    \caption{Sensitivity of the proposed experiment to the axion-photon coupling coefficient for time integration up to the ALPs coherence time (black solid line) and using twin polarimeters with a one-year integration time (red dashed). We evaluated the sensitivity of the proposed polarimetry experiment with two quarter-wave plates near the mirrors of a 5-meter-long cavity, with 20 kW circulating power and limited by shot-noise. For comparison, the existing constraints from CAST~\cite{anastassopoulos2017new} (blue) and the design sensitivity for ALPS~II~\cite{Bahre:2013ywa} (green) and the quantum-enhanced interferometry detector~\cite{martynov_quantum-enhanced_2020} (dotted dashed) are shown.}
    \label{fig:axion_prospects}
\end{figure}

To calculate the sensitivity to Axion DM, we must estimate the sensitivity of a polarimeter including two thin quarter-wave plates inside a Fabry-Pérot cavity. A quarter wave plate, constructed as a zero-order wave plate (where the relative phase retardation is $\pi/2$ rather than an integer multiple thereof) can be very thin, so bulk absorption can be assumed to be less than 10~ppm. Thus, the losses caused by the AR coating of the two wave plates will be the most significant effect. With 25~ppm loss per surface, two quarter-wave plates will contribute to 100~ppm loss per trip. Given that the losses exceed the mirrors' transmittance $P\gg T$, the losses that occur in the cavity are $P=P_{\rm AR}+\mu d\approx P_{\rm AR}$. The sensitivity as described by Eq.~\ref{eq:signal_dem_pemADM} is evaluated as follows:

\begin{eqnarray}
 |\rho(\nu)| &=&  \frac{S_{\rm P}^{\rm (tot)}}{N h_{\rm T}(\nu)}\\
\nonumber &\approx& S_{\rm P}^{\rm (tot)} \frac{\sqrt{P_{\rm AR}^2+4 \sin^2{\pi \nu \tau}}}{2}~\text{for}~ T+P^*<P_{\rm AR}
    \label{eq:axion}
\end{eqnarray}

where $P^*$ is the sum of all the other absorption in the polarimeter. Considering Eq.~\ref{eq:axion_osc}, the projected sensitivity in terms of the axion-photon coupling constant as a function of the axion frequency is 

\begin{equation}
    g_{\alpha\gamma}=\frac{S_{\rm P}^{\rm (tot)}}{2 \tau}\sqrt{\frac{P_{\rm AR}^2+4 \sin ^2\left( \pi  \nu_{\rm a} \tau\right)}{2 \rho_\mathrm{local}}},
\end{equation}

where $\tau$ is the cavity round-trip travel time. The Fig.~\ref{fig:axion_prospects} shows the polarimeter sensitivity for an integration time up to the coherence time of the axion ($t_{\rm int}=10^6/\nu_{\rm a}$), and using twin polarimeters with a one-year cross-correlation time for two identical polarimeters with 5-meter cavities, each configured as shown in Fig.~\ref{scheme-newADM}. Existing constraints from the solar CERN Axion Solar Telescope (CAST) \cite{anastassopoulos2017new}, the design sensitivity for Any Light Particle Search (ALPS\,II) \cite{Bahre:2013ywa}, and the quantum-enhanced interferometry detector \cite{martynov_quantum-enhanced_2020} are shown for comparison.

\section{Discussion and Conclusion}
\label{sec:discussion}

We propose the use of polarimetry for direct searches for axion-like particles and for low-mass scalar-field dark matter. Projections of realistic sensitivities of the proposed polarimetry techniques were made and it was found that such experiments would be more sensitive than existing experiments for a wide range of possible DM particle masses. 

We showed that scalar field DM interacting with a highly birefringent crystal (yttrium vanadate or sapphire) drives thickness and refractive index variation of the crystal at the frequency of the scalar field. This, in turn, produces differential phase oscillations between orthogonal polarization components of light traversing the crystal. If the birefringent crystal is placed in a Fabry-Pérot cavity, these differential phase oscillations between polarization components can be measured with a polarimeter at high sensitivity using a heterodyne readout technique. The amplification of the signal affected by the Fabry-Pérot cavity is limited by absorption in the birefringent crystal. Regardless, with a 5~cm-thick birefringent yttrium vanadate crystal, the polarimetry instrument could probe an unexplored region of the scalar field DM parameter space at DM masses ranging from $\left(10^{-15}-10^{-9}\right)$~eV and frequencies of $1$~Hz to $200$~kHz, beyond existing constraints \cite{vermeulen2021direct, aiello_constraints_2021, tretiak_improved_2022}. As the sensitivity scales linearly with the solid bulk absorption, potential technological advances that allow the production of purer crystals with less bulk absorption could significantly improve the sensitivity of the proposed polarimetry method. 
 
The same polarimetry setup, reconfigured with two quarter-wave plates placed near the mirrors instead of a thick birefringent solid, could be used to search for axion-like particles with masses ranging from $\left(10^{-15}-10^{-8}\right)$~eV and frequencies ranging from 1~Hz to 2.5~MHz. Axion-like particles may be present in a galactic halo; if so, they will manifest themselves by annihilating and producing photons in a polarimeter. This would result in a phase oscillation between the orthogonal polarizations of light. For the axion-like field, the sensitivity will be limited by the AR coating of the QWPs, which will limit the amplification factor of the Fabry-Pérot cavity. Even a 5~m-cavity polarimeter, easily made as a tabletop prototype, has the potential to surpass the sensitivity of  CAST~\cite{anastassopoulos2017new} in the ALPs mass range of $10^{-15}$~eV all the way up to $10^{-9}$~eV. The sensitivity can be improved by using a longer cavity. This would improve the signal-to-noise ratio at low frequencies but would limit the sensitive bandwidth, thus decreasing the sensitivity to high axion masses.

Moreover, existing gravitational wave facilities might be modified to function as kilometer-scale polarimeters for DM searches similar to the one proposed in \cite{derocco_axion_2018}. The polarimeter setup would be easy to integrate into GW detectors with minimal changes, and the sensitivity would be five to seven orders of magnitude better than the CAST constraints. The current gravitational-wave facilities would offer great infrastructure for the building of axion polarimetry as they already possess vacuum systems, cutting-edge suspensions, and powerful lasers.  The third-generation gravitational-wave facilities would then not only be able to detect GWs from distant regions of the cosmos \cite{Punturo_2010, Abbott_2017}, but would also enable a potential detection of dark matter using the same facility.

In both instances, at sub-Hertz frequencies, we estimate that the sensitivity restrictions will mostly be dominated by seismic and $1/f$ noises rather than quantum noises. 
We must also consider the effect of possible spurious birefringence, which could arise due to e.g. birefringence of the mirrors. Compensating for static birefringence caused by the Fabry-P\`erot mirrors is feasible by rotating the axis of the birefringent medium. Effective minimization of the effect of birefringent imperfections in both quarter wave plates and birefringent solids can be achieved, as exemplified in the research on polarimetry by Zavattini et al. (2022) \cite{zavattini2022polarimetry}. Furthermore, while wide band birefringence noise caused by the mirrors has the potential to limit sensitivity, it has been found to be directly proportional to the cavity build-up factor and is dominant only at high build-up factors as reported \cite{zavattini2022polarimetry}, which is above typical build-up levels consider in this paper. Importantly, for both of the dark matter scenarios considered, our proposal is to use cross-correlation, which would have these detectors look for the expected coherent spatial harmonic oscillation modes of dark matter candidates.

\section*{Acknowledgments}

A.E. wants to thank G. Zavattini, F. Della Valle, G. Messineo, E. Mariotti, and S. Kunc for the useful conversation about the suggested scalar DM experiment. L.A. would like to thank Daniel Brown for useful discussions about thermal effects computation in FINESSE 3. We thank the support of the Leverhulme Trust in the United Kingdom via grants RPG-2019-022 and PLP-2018-066 and Science and Technology Facilities Council (STFC) via grants ST/T006331/1 and ST/W6456/1.

\bibliographystyle{apsrev4-2}
\bibliography{bibliography.bib}

\begin{thebibliography}{54}%
\makeatletter
\providecommand \@ifxundefined [1]{%
 \@ifx{#1\undefined}
}%
\providecommand \@ifnum [1]{%
 \ifnum #1\expandafter \@firstoftwo
 \else \expandafter \@secondoftwo
 \fi
}%
\providecommand \@ifx [1]{%
 \ifx #1\expandafter \@firstoftwo
 \else \expandafter \@secondoftwo
 \fi
}%
\providecommand \natexlab [1]{#1}%
\providecommand \enquote  [1]{``#1''}%
\providecommand \bibnamefont  [1]{#1}%
\providecommand \bibfnamefont [1]{#1}%
\providecommand \citenamefont [1]{#1}%
\providecommand \href@noop [0]{\@secondoftwo}%
\providecommand \href [0]{\begingroup \@sanitize@url \@href}%
\providecommand \@href[1]{\@@startlink{#1}\@@href}%
\providecommand \@@href[1]{\endgroup#1\@@endlink}%
\providecommand \@sanitize@url [0]{\catcode `\\12\catcode `\$12\catcode
  `\&12\catcode `\#12\catcode `\^12\catcode `\_12\catcode `\%12\relax}%
\providecommand \@@startlink[1]{}%
\providecommand \@@endlink[0]{}%
\providecommand \url  [0]{\begingroup\@sanitize@url \@url }%
\providecommand \@url [1]{\endgroup\@href {#1}{\urlprefix }}%
\providecommand \urlprefix  [0]{URL }%
\providecommand \Eprint [0]{\href }%
\providecommand \doibase [0]{https://doi.org/}%
\providecommand \selectlanguage [0]{\@gobble}%
\providecommand \bibinfo  [0]{\@secondoftwo}%
\providecommand \bibfield  [0]{\@secondoftwo}%
\providecommand \translation [1]{[#1]}%
\providecommand \BibitemOpen [0]{}%
\providecommand \bibitemStop [0]{}%
\providecommand \bibitemNoStop [0]{.\EOS\space}%
\providecommand \EOS [0]{\spacefactor3000\relax}%
\providecommand \BibitemShut  [1]{\csname bibitem#1\endcsname}%
\let\auto@bib@innerbib\@empty
\bibitem [{\citenamefont {Aghanim}\ \emph {et~al.}(2020)\citenamefont
  {Aghanim}, \citenamefont {Akrami}, \citenamefont {Ashdown}, \citenamefont
  {Aumont}, \citenamefont {Baccigalupi}, \citenamefont {Ballardini},
  \citenamefont {Banday}, \citenamefont {Barreiro}, \citenamefont {Bartolo},
  \citenamefont {Basak} \emph {et~al.}}]{aghanim2020planck}%
  \BibitemOpen
  \bibfield  {author} {\bibinfo {author} {\bibfnamefont {N.}~\bibnamefont
  {Aghanim}}, \bibinfo {author} {\bibfnamefont {Y.}~\bibnamefont {Akrami}},
  \bibinfo {author} {\bibfnamefont {M.}~\bibnamefont {Ashdown}}, \bibinfo
  {author} {\bibfnamefont {J.}~\bibnamefont {Aumont}}, \bibinfo {author}
  {\bibfnamefont {C.}~\bibnamefont {Baccigalupi}}, \bibinfo {author}
  {\bibfnamefont {M.}~\bibnamefont {Ballardini}}, \bibinfo {author}
  {\bibfnamefont {A.}~\bibnamefont {Banday}}, \bibinfo {author} {\bibfnamefont
  {R.}~\bibnamefont {Barreiro}}, \bibinfo {author} {\bibfnamefont
  {N.}~\bibnamefont {Bartolo}}, \bibinfo {author} {\bibfnamefont
  {S.}~\bibnamefont {Basak}}, \emph {et~al.},\ }\href
  {https://doi.org/10.1051/0004-6361/201833910} {\bibfield  {journal} {\bibinfo
   {journal} {Astronomy \& Astrophysics}\ }\textbf {\bibinfo {volume} {641}},\
  \bibinfo {pages} {A6} (\bibinfo {year} {2020})}\BibitemShut {NoStop}%
\bibitem [{\citenamefont {Aprile}\ \emph {et~al.}(2017)\citenamefont {Aprile},
  \citenamefont {Aalbers}, \citenamefont {Agostini}, \citenamefont {Alfonsi},
  \citenamefont {Amaro}, \citenamefont {Anthony}, \citenamefont {Arneodo},
  \citenamefont {Barrow}, \citenamefont {Baudis}, \citenamefont {Bauermeister}
  \emph {et~al.}}]{aprile2017xenon1t}%
  \BibitemOpen
  \bibfield  {author} {\bibinfo {author} {\bibfnamefont {E.}~\bibnamefont
  {Aprile}}, \bibinfo {author} {\bibfnamefont {J.}~\bibnamefont {Aalbers}},
  \bibinfo {author} {\bibfnamefont {F.}~\bibnamefont {Agostini}}, \bibinfo
  {author} {\bibfnamefont {M.}~\bibnamefont {Alfonsi}}, \bibinfo {author}
  {\bibfnamefont {F.}~\bibnamefont {Amaro}}, \bibinfo {author} {\bibfnamefont
  {M.}~\bibnamefont {Anthony}}, \bibinfo {author} {\bibfnamefont
  {F.}~\bibnamefont {Arneodo}}, \bibinfo {author} {\bibfnamefont
  {P.}~\bibnamefont {Barrow}}, \bibinfo {author} {\bibfnamefont
  {L.}~\bibnamefont {Baudis}}, \bibinfo {author} {\bibfnamefont
  {B.}~\bibnamefont {Bauermeister}}, \emph {et~al.},\ }\href
  {https://doi.org/10.1140/epjc/s10052-017-4902-x} {\bibfield  {journal}
  {\bibinfo  {journal} {Eur. Phys. J. C}\ }\textbf {\bibinfo {volume} {77}},\
  \bibinfo {pages} {1} (\bibinfo {year} {2017})}\BibitemShut {NoStop}%
\bibitem [{\citenamefont {Akerib}\ \emph {et~al.}(2013)\citenamefont {Akerib},
  \citenamefont {Bai}, \citenamefont {Bedikian}, \citenamefont {Bernard},
  \citenamefont {Bernstein}, \citenamefont {Bolozdynya}, \citenamefont
  {Bradley}, \citenamefont {Byram}, \citenamefont {Cahn}, \citenamefont {Camp}
  \emph {et~al.}}]{akerib2013large}%
  \BibitemOpen
  \bibfield  {author} {\bibinfo {author} {\bibfnamefont {D.}~\bibnamefont
  {Akerib}}, \bibinfo {author} {\bibfnamefont {X.}~\bibnamefont {Bai}},
  \bibinfo {author} {\bibfnamefont {S.}~\bibnamefont {Bedikian}}, \bibinfo
  {author} {\bibfnamefont {E.}~\bibnamefont {Bernard}}, \bibinfo {author}
  {\bibfnamefont {A.}~\bibnamefont {Bernstein}}, \bibinfo {author}
  {\bibfnamefont {A.}~\bibnamefont {Bolozdynya}}, \bibinfo {author}
  {\bibfnamefont {A.}~\bibnamefont {Bradley}}, \bibinfo {author} {\bibfnamefont
  {D.}~\bibnamefont {Byram}}, \bibinfo {author} {\bibfnamefont
  {S.}~\bibnamefont {Cahn}}, \bibinfo {author} {\bibfnamefont {C.}~\bibnamefont
  {Camp}}, \emph {et~al.},\ }\href {https://doi.org/10.1016/j.nima.2012.11.135}
  {\bibfield  {journal} {\bibinfo  {journal} {Nucl. Instrum. Methods Phys.
  Res., Sect. A}\ }\textbf {\bibinfo {volume} {704}},\ \bibinfo {pages} {111}
  (\bibinfo {year} {2013})}\BibitemShut {NoStop}%
\bibitem [{\citenamefont {Zhang}\ \emph {et~al.}(2019)\citenamefont {Zhang},
  \citenamefont {Abdukerim}, \citenamefont {Chen}, \citenamefont {Chen},
  \citenamefont {Chen}, \citenamefont {Cui}, \citenamefont {Dong},
  \citenamefont {Fang}, \citenamefont {Fu}, \citenamefont {Giboni} \emph
  {et~al.}}]{zhang2019dark}%
  \BibitemOpen
  \bibfield  {author} {\bibinfo {author} {\bibfnamefont {H.}~\bibnamefont
  {Zhang}}, \bibinfo {author} {\bibfnamefont {A.}~\bibnamefont {Abdukerim}},
  \bibinfo {author} {\bibfnamefont {W.}~\bibnamefont {Chen}}, \bibinfo {author}
  {\bibfnamefont {X.}~\bibnamefont {Chen}}, \bibinfo {author} {\bibfnamefont
  {Y.}~\bibnamefont {Chen}}, \bibinfo {author} {\bibfnamefont {X.}~\bibnamefont
  {Cui}}, \bibinfo {author} {\bibfnamefont {B.}~\bibnamefont {Dong}}, \bibinfo
  {author} {\bibfnamefont {D.}~\bibnamefont {Fang}}, \bibinfo {author}
  {\bibfnamefont {C.}~\bibnamefont {Fu}}, \bibinfo {author} {\bibfnamefont
  {K.}~\bibnamefont {Giboni}}, \emph {et~al.},\ }\href
  {https://doi.org/10.1007/s11433-018-9259-0} {\bibfield  {journal} {\bibinfo
  {journal} {Science China Physics, Mechanics \& Astronomy}\ }\textbf {\bibinfo
  {volume} {62}},\ \bibinfo {pages} {1} (\bibinfo {year} {2019})}\BibitemShut
  {NoStop}%
\bibitem [{\citenamefont {Baudis}\ \emph {et~al.}(2014)\citenamefont {Baudis},
  \citenamefont {Ferella}, \citenamefont {Kish}, \citenamefont {Manalaysay},
  \citenamefont {Undagoitia},\ and\ \citenamefont
  {Schumann}}]{baudis2014neutrino}%
  \BibitemOpen
  \bibfield  {author} {\bibinfo {author} {\bibfnamefont {L.}~\bibnamefont
  {Baudis}}, \bibinfo {author} {\bibfnamefont {A.}~\bibnamefont {Ferella}},
  \bibinfo {author} {\bibfnamefont {A.}~\bibnamefont {Kish}}, \bibinfo {author}
  {\bibfnamefont {A.}~\bibnamefont {Manalaysay}}, \bibinfo {author}
  {\bibfnamefont {T.~M.}\ \bibnamefont {Undagoitia}},\ and\ \bibinfo {author}
  {\bibfnamefont {M.}~\bibnamefont {Schumann}},\ }\href
  {https://doi.org/10.1088/1475-7516/2014/01/044} {\bibfield  {journal}
  {\bibinfo  {journal} {J. Cosmol. Astropart. Phys.}\ }\textbf {\bibinfo
  {volume} {2014}}\bibinfo  {number} { (01)},\ \bibinfo {pages}
  {044}}\BibitemShut {NoStop}%
\bibitem [{\citenamefont {Canepa}(2019)}]{canepa2019searches}%
  \BibitemOpen
\bibfield  {number} {  }\bibfield  {author} {\bibinfo {author} {\bibfnamefont
  {A.}~\bibnamefont {Canepa}},\ }\href
  {https://doi.org/https://doi.org/10.1016/j.revip.2019.100033} {\bibfield
  {journal} {\bibinfo  {journal} {Rev. Phys.}\ }\textbf {\bibinfo {volume}
  {4}},\ \bibinfo {pages} {100033} (\bibinfo {year} {2019})}\BibitemShut
  {NoStop}%
\bibitem [{\citenamefont {Stadnik}\ and\ \citenamefont
  {Flambaum}(2015{\natexlab{a}})}]{stadnik2015searching}%
  \BibitemOpen
  \bibfield  {author} {\bibinfo {author} {\bibfnamefont {Y.~V.}\ \bibnamefont
  {Stadnik}}\ and\ \bibinfo {author} {\bibfnamefont {V.~V.}\ \bibnamefont
  {Flambaum}},\ }\href {https://doi.org/10.1103/PhysRevLett.114.161301}
  {\bibfield  {journal} {\bibinfo  {journal} {Phys. Rev. Lett.}\ }\textbf
  {\bibinfo {volume} {114}},\ \bibinfo {pages} {161301} (\bibinfo {year}
  {2015}{\natexlab{a}})}\BibitemShut {NoStop}%
\bibitem [{\citenamefont {Stadnik}\ and\ \citenamefont
  {Flambaum}(2015{\natexlab{b}})}]{stadnik2015can}%
  \BibitemOpen
  \bibfield  {author} {\bibinfo {author} {\bibfnamefont {Y.~V.}\ \bibnamefont
  {Stadnik}}\ and\ \bibinfo {author} {\bibfnamefont {V.~V.}\ \bibnamefont
  {Flambaum}},\ }\href {https://doi.org/10.1103/PhysRevLett.115.201301}
  {\bibfield  {journal} {\bibinfo  {journal} {Phys. Rev. Lett.}\ }\textbf
  {\bibinfo {volume} {115}},\ \bibinfo {pages} {201301} (\bibinfo {year}
  {2015}{\natexlab{b}})}\BibitemShut {NoStop}%
\bibitem [{\citenamefont {Graham}\ and\ \citenamefont
  {Rajendran}(2013)}]{PhysRevD.88.035023}%
  \BibitemOpen
  \bibfield  {author} {\bibinfo {author} {\bibfnamefont {P.}~\bibnamefont
  {Graham}}\ and\ \bibinfo {author} {\bibfnamefont {S.}~\bibnamefont
  {Rajendran}},\ }\href {https://doi.org/10.1103/PhysRevD.88.035023} {\bibfield
   {journal} {\bibinfo  {journal} {Phys. Rev. D}\ }\textbf {\bibinfo {volume}
  {88}},\ \bibinfo {pages} {035023} (\bibinfo {year} {2013})}\BibitemShut
  {NoStop}%
\bibitem [{\citenamefont {Ringwald}(2014)}]{Ringwald_2014}%
  \BibitemOpen
  \bibfield  {author} {\bibinfo {author} {\bibfnamefont {A.}~\bibnamefont
  {Ringwald}},\ }\href {https://doi.org/10.1088/1742-6596/485/1/012013}
  {\bibfield  {journal} {\bibinfo  {journal} {J. Phys. Conf. Ser.}\ }\textbf
  {\bibinfo {volume} {485}},\ \bibinfo {pages} {012013} (\bibinfo {year}
  {2014})}\BibitemShut {NoStop}%
\bibitem [{\citenamefont {Farina}\ \emph {et~al.}(2017)\citenamefont {Farina},
  \citenamefont {Pappadopulo}, \citenamefont {Rompineve},\ and\ \citenamefont
  {Tesi}}]{farina2017photo}%
  \BibitemOpen
  \bibfield  {author} {\bibinfo {author} {\bibfnamefont {M.}~\bibnamefont
  {Farina}}, \bibinfo {author} {\bibfnamefont {D.}~\bibnamefont {Pappadopulo}},
  \bibinfo {author} {\bibfnamefont {F.}~\bibnamefont {Rompineve}},\ and\
  \bibinfo {author} {\bibfnamefont {A.}~\bibnamefont {Tesi}},\ }\href
  {https://doi.org/10.1007/JHEP01(2017)095} {\bibfield  {journal} {\bibinfo
  {journal} {J. High Energy Phys.}\ }\textbf {\bibinfo {volume} {2017}}\bibinfo
   {number} { (001)},\ \bibinfo {pages} {1}}\BibitemShut {NoStop}%
\bibitem [{\citenamefont {Nagano}\ \emph {et~al.}(2021)\citenamefont {Nagano},
  \citenamefont {Nakatsuka}, \citenamefont {Morisaki}, \citenamefont {Fujita},
  \citenamefont {Michimura},\ and\ \citenamefont
  {Obata}}]{PhysRevD.104.062008}%
  \BibitemOpen
\bibfield  {number} {  }\bibfield  {author} {\bibinfo {author} {\bibfnamefont
  {K.}~\bibnamefont {Nagano}}, \bibinfo {author} {\bibfnamefont
  {H.}~\bibnamefont {Nakatsuka}}, \bibinfo {author} {\bibfnamefont
  {S.}~\bibnamefont {Morisaki}}, \bibinfo {author} {\bibfnamefont
  {T.}~\bibnamefont {Fujita}}, \bibinfo {author} {\bibfnamefont
  {Y.}~\bibnamefont {Michimura}},\ and\ \bibinfo {author} {\bibfnamefont
  {I.}~\bibnamefont {Obata}},\ }\href
  {https://doi.org/10.1103/PhysRevD.104.062008} {\bibfield  {journal} {\bibinfo
   {journal} {Phys. Rev. D}\ }\textbf {\bibinfo {volume} {104}},\ \bibinfo
  {pages} {062008} (\bibinfo {year} {2021})}\BibitemShut {NoStop}%
\bibitem [{\citenamefont {Grote}\ and\ \citenamefont
  {Stadnik}(2019)}]{grote_novel_2019}%
  \BibitemOpen
  \bibfield  {author} {\bibinfo {author} {\bibfnamefont {H.}~\bibnamefont
  {Grote}}\ and\ \bibinfo {author} {\bibfnamefont {Y.~V.}\ \bibnamefont
  {Stadnik}},\ }\href {https://doi.org/10.1103/PhysRevResearch.1.033187}
  {\bibfield  {journal} {\bibinfo  {journal} {Phys. Rev. Res.}\ }\textbf
  {\bibinfo {volume} {1}},\ \bibinfo {pages} {033187} (\bibinfo {year}
  {2019})}\BibitemShut {NoStop}%
\bibitem [{\citenamefont {Vermeulen}\ \emph
  {et~al.}(2021{\natexlab{a}})\citenamefont {Vermeulen}, \citenamefont
  {Relton}, \citenamefont {Grote}, \citenamefont {Raymond}, \citenamefont
  {Affeldt}, \citenamefont {Bergamin}, \citenamefont {Bisht}, \citenamefont
  {Brinkmann}, \citenamefont {Danzmann}, \citenamefont {Doravari} \emph
  {et~al.}}]{vermeulen2021direct}%
  \BibitemOpen
  \bibfield  {author} {\bibinfo {author} {\bibfnamefont {S.}~\bibnamefont
  {Vermeulen}}, \bibinfo {author} {\bibfnamefont {P.}~\bibnamefont {Relton}},
  \bibinfo {author} {\bibfnamefont {H.}~\bibnamefont {Grote}}, \bibinfo
  {author} {\bibfnamefont {V.}~\bibnamefont {Raymond}}, \bibinfo {author}
  {\bibfnamefont {C.}~\bibnamefont {Affeldt}}, \bibinfo {author} {\bibfnamefont
  {F.}~\bibnamefont {Bergamin}}, \bibinfo {author} {\bibfnamefont
  {A.}~\bibnamefont {Bisht}}, \bibinfo {author} {\bibfnamefont
  {M.}~\bibnamefont {Brinkmann}}, \bibinfo {author} {\bibfnamefont
  {K.}~\bibnamefont {Danzmann}}, \bibinfo {author} {\bibfnamefont
  {S.}~\bibnamefont {Doravari}}, \emph {et~al.},\ }\href
  {https://doi.org/10.1038/s41586-021-04031-y} {\bibfield  {journal} {\bibinfo
  {journal} {Nature}\ }\textbf {\bibinfo {volume} {600}},\ \bibinfo {pages}
  {424} (\bibinfo {year} {2021}{\natexlab{a}})}\BibitemShut {NoStop}%
\bibitem [{\citenamefont {Aiello}\ \emph {et~al.}(2022)\citenamefont {Aiello},
  \citenamefont {Richardson}, \citenamefont {Vermeulen}, \citenamefont {Grote},
  \citenamefont {Hogan}, \citenamefont {Kwon},\ and\ \citenamefont
  {Stoughton}}]{aiello_constraints_2021}%
  \BibitemOpen
  \bibfield  {author} {\bibinfo {author} {\bibfnamefont {L.}~\bibnamefont
  {Aiello}}, \bibinfo {author} {\bibfnamefont {J.}~\bibnamefont {Richardson}},
  \bibinfo {author} {\bibfnamefont {S.}~\bibnamefont {Vermeulen}}, \bibinfo
  {author} {\bibfnamefont {H.}~\bibnamefont {Grote}}, \bibinfo {author}
  {\bibfnamefont {C.}~\bibnamefont {Hogan}}, \bibinfo {author} {\bibfnamefont
  {O.}~\bibnamefont {Kwon}},\ and\ \bibinfo {author} {\bibfnamefont
  {C.}~\bibnamefont {Stoughton}},\ }\href
  {https://doi.org/10.1103/PhysRevLett.128.121101} {\bibfield  {journal}
  {\bibinfo  {journal} {Phys. Rev. Lett.}\ }\textbf {\bibinfo {volume} {128}},\
  \bibinfo {pages} {121101} (\bibinfo {year} {2022})}\BibitemShut {NoStop}%
\bibitem [{\citenamefont {Vermeulen}\ \emph
  {et~al.}(2021{\natexlab{b}})\citenamefont {Vermeulen}, \citenamefont
  {Aiello}, \citenamefont {Ejlli}, \citenamefont {Griffiths}, \citenamefont
  {James}, \citenamefont {Dooley},\ and\ \citenamefont
  {Grote}}]{vermeulen2021experiment}%
  \BibitemOpen
  \bibfield  {author} {\bibinfo {author} {\bibfnamefont {S.}~\bibnamefont
  {Vermeulen}}, \bibinfo {author} {\bibfnamefont {L.}~\bibnamefont {Aiello}},
  \bibinfo {author} {\bibfnamefont {A.}~\bibnamefont {Ejlli}}, \bibinfo
  {author} {\bibfnamefont {W.}~\bibnamefont {Griffiths}}, \bibinfo {author}
  {\bibfnamefont {A.}~\bibnamefont {James}}, \bibinfo {author} {\bibfnamefont
  {K.}~\bibnamefont {Dooley}},\ and\ \bibinfo {author} {\bibfnamefont
  {H.}~\bibnamefont {Grote}},\ }\href
  {https://doi.org/10.1088/1361-6382/abe757} {\bibfield  {journal} {\bibinfo
  {journal} {Classical Quantum Gravity}\ }\textbf {\bibinfo {volume} {38}},\
  \bibinfo {pages} {085008} (\bibinfo {year} {2021}{\natexlab{b}})}\BibitemShut
  {NoStop}%
\bibitem [{\citenamefont {Martynov}\ and\ \citenamefont
  {Miao}(2020)}]{martynov_quantum-enhanced_2020}%
  \BibitemOpen
  \bibfield  {author} {\bibinfo {author} {\bibfnamefont {D.}~\bibnamefont
  {Martynov}}\ and\ \bibinfo {author} {\bibfnamefont {H.}~\bibnamefont
  {Miao}},\ }\href {https://doi.org/10.1103/PhysRevD.101.095034} {\bibfield
  {journal} {\bibinfo  {journal} {Phys. Rev. D}\ }\textbf {\bibinfo {volume}
  {101}},\ \bibinfo {pages} {095034} (\bibinfo {year} {2020})}\BibitemShut
  {NoStop}%
\bibitem [{\citenamefont {Euler}\ and\ \citenamefont
  {Kockel}(1935)}]{euler_uber_1935}%
  \BibitemOpen
  \bibfield  {author} {\bibinfo {author} {\bibfnamefont {H.}~\bibnamefont
  {Euler}}\ and\ \bibinfo {author} {\bibfnamefont {B.}~\bibnamefont {Kockel}},\
  }\href {https://doi.org/10.1007/BF01493898} {\bibfield  {journal} {\bibinfo
  {journal} {Naturwissenschaften}\ }\textbf {\bibinfo {volume} {23}},\ \bibinfo
  {pages} {246} (\bibinfo {year} {1935})}\BibitemShut {NoStop}%
\bibitem [{\citenamefont {Erber}(1961)}]{erber_velocity_1961}%
  \BibitemOpen
  \bibfield  {author} {\bibinfo {author} {\bibfnamefont {T.}~\bibnamefont
  {Erber}},\ }\href {https://doi.org/10.1038/190025a0} {\bibfield  {journal}
  {\bibinfo  {journal} {Nature}\ }\textbf {\bibinfo {volume} {190}},\ \bibinfo
  {pages} {25} (\bibinfo {year} {1961})}\BibitemShut {NoStop}%
\bibitem [{\citenamefont {Iacopini}\ and\ \citenamefont
  {Zavattini}(1979)}]{iacopini_experimental_1979}%
  \BibitemOpen
  \bibfield  {author} {\bibinfo {author} {\bibfnamefont {E.}~\bibnamefont
  {Iacopini}}\ and\ \bibinfo {author} {\bibfnamefont {E.}~\bibnamefont
  {Zavattini}},\ }\href {https://doi.org/10.1016/0370-2693(79)90797-4}
  {\bibfield  {journal} {\bibinfo  {journal} {Phys. Lett. B}\ }\textbf
  {\bibinfo {volume} {85}},\ \bibinfo {pages} {151} (\bibinfo {year}
  {1979})}\BibitemShut {NoStop}%
\bibitem [{\citenamefont {Zavattini}\ \emph {et~al.}(2022)\citenamefont
  {Zavattini}, \citenamefont {Della~Valle}, \citenamefont {Soflau},
  \citenamefont {Formaggio}, \citenamefont {Crapulli}, \citenamefont
  {Messineo}, \citenamefont {Mariotti}, \citenamefont {Kunc}, \citenamefont
  {Ejlli}, \citenamefont {Ruoso}, \citenamefont {Marinelli},\ and\
  \citenamefont {Andreotti}}]{zavattini2022polarimetry}%
  \BibitemOpen
  \bibfield  {author} {\bibinfo {author} {\bibfnamefont {G.}~\bibnamefont
  {Zavattini}}, \bibinfo {author} {\bibfnamefont {F.}~\bibnamefont
  {Della~Valle}}, \bibinfo {author} {\bibfnamefont {A.}~\bibnamefont {Soflau}},
  \bibinfo {author} {\bibfnamefont {L.}~\bibnamefont {Formaggio}}, \bibinfo
  {author} {\bibfnamefont {G.}~\bibnamefont {Crapulli}}, \bibinfo {author}
  {\bibfnamefont {G.}~\bibnamefont {Messineo}}, \bibinfo {author}
  {\bibfnamefont {E.}~\bibnamefont {Mariotti}}, \bibinfo {author}
  {\bibfnamefont {S.}~\bibnamefont {Kunc}}, \bibinfo {author} {\bibfnamefont
  {A.}~\bibnamefont {Ejlli}}, \bibinfo {author} {\bibfnamefont
  {G.}~\bibnamefont {Ruoso}}, \bibinfo {author} {\bibfnamefont
  {C.}~\bibnamefont {Marinelli}},\ and\ \bibinfo {author} {\bibfnamefont
  {M.}~\bibnamefont {Andreotti}},\ }\href
  {https://doi.org/10.1140/epjc/s10052-016-4139-0} {\bibfield  {journal}
  {\bibinfo  {journal} {Eur. Phys. J. C}\ }\textbf {\bibinfo {volume} {82}},\
  \bibinfo {pages} {159} (\bibinfo {year} {2022})}\BibitemShut {NoStop}%
\bibitem [{\citenamefont {Ejlli}\ \emph {et~al.}(2020)\citenamefont {Ejlli},
  \citenamefont {Della~Valle}, \citenamefont {Gastaldi}, \citenamefont
  {Messineo}, \citenamefont {Pengo}, \citenamefont {Ruoso},\ and\ \citenamefont
  {Zavattini}}]{ejlli_pvlas_2020}%
  \BibitemOpen
  \bibfield  {author} {\bibinfo {author} {\bibfnamefont {A.}~\bibnamefont
  {Ejlli}}, \bibinfo {author} {\bibfnamefont {F.}~\bibnamefont {Della~Valle}},
  \bibinfo {author} {\bibfnamefont {U.}~\bibnamefont {Gastaldi}}, \bibinfo
  {author} {\bibfnamefont {G.}~\bibnamefont {Messineo}}, \bibinfo {author}
  {\bibfnamefont {R.}~\bibnamefont {Pengo}}, \bibinfo {author} {\bibfnamefont
  {G.}~\bibnamefont {Ruoso}},\ and\ \bibinfo {author} {\bibfnamefont
  {G.}~\bibnamefont {Zavattini}},\ }\href
  {https://doi.org/10.1016/j.physrep.2020.06.001} {\bibfield  {journal}
  {\bibinfo  {journal} {Phys. Rep.}\ }\textbf {\bibinfo {volume} {871}},\
  \bibinfo {pages} {1} (\bibinfo {year} {2020})}\BibitemShut {NoStop}%
\bibitem [{\citenamefont {Derevianko}(2018)}]{derevianko_detecting_2018}%
  \BibitemOpen
  \bibfield  {author} {\bibinfo {author} {\bibfnamefont {A.}~\bibnamefont
  {Derevianko}},\ }\href {https://doi.org/10.1103/PhysRevA.97.042506}
  {\bibfield  {journal} {\bibinfo  {journal} {Phys. Rev. A}\ }\textbf {\bibinfo
  {volume} {97}},\ \bibinfo {pages} {042506} (\bibinfo {year}
  {2018})}\BibitemShut {NoStop}%
\bibitem [{\citenamefont {Arvanitaki}\ \emph {et~al.}(2015)\citenamefont
  {Arvanitaki}, \citenamefont {Huang},\ and\ \citenamefont
  {Van~Tilburg}}]{arvanitaki_searching_2015}%
  \BibitemOpen
  \bibfield  {author} {\bibinfo {author} {\bibfnamefont {A.}~\bibnamefont
  {Arvanitaki}}, \bibinfo {author} {\bibfnamefont {J.}~\bibnamefont {Huang}},\
  and\ \bibinfo {author} {\bibfnamefont {K.}~\bibnamefont {Van~Tilburg}},\
  }\href {https://doi.org/10.1103/PhysRevD.91.015015} {\bibfield  {journal}
  {\bibinfo  {journal} {Phys. Rev. D}\ }\textbf {\bibinfo {volume} {91}},\
  \bibinfo {pages} {015015} (\bibinfo {year} {2015})}\BibitemShut {NoStop}%
\bibitem [{\citenamefont {Sol\'{\i}s-L\'opez}\ \emph
  {et~al.}(2021)\citenamefont {Sol\'{\i}s-L\'opez}, \citenamefont {Guzm\'an},
  \citenamefont {Matos}, \citenamefont {Robles},\ and\ \citenamefont {Ure\~na
  L\'opez}}]{PhysRevD.103.083535}%
  \BibitemOpen
  \bibfield  {author} {\bibinfo {author} {\bibfnamefont {J.}~\bibnamefont
  {Sol\'{\i}s-L\'opez}}, \bibinfo {author} {\bibfnamefont {F.~S.}\ \bibnamefont
  {Guzm\'an}}, \bibinfo {author} {\bibfnamefont {T.}~\bibnamefont {Matos}},
  \bibinfo {author} {\bibfnamefont {V.}~\bibnamefont {Robles}},\ and\ \bibinfo
  {author} {\bibfnamefont {L.~A.}\ \bibnamefont {Ure\~na L\'opez}},\ }\href
  {https://doi.org/10.1103/PhysRevD.103.083535} {\bibfield  {journal} {\bibinfo
   {journal} {Phys. Rev. D}\ }\textbf {\bibinfo {volume} {103}},\ \bibinfo
  {pages} {083535} (\bibinfo {year} {2021})}\BibitemShut {NoStop}%
\bibitem [{\citenamefont {Aprile}\ \emph {et~al.}(2021)\citenamefont {Aprile}
  \emph {et~al.}}]{XENON:2020fgj}%
  \BibitemOpen
  \bibfield  {author} {\bibinfo {author} {\bibfnamefont {E.~X.~C.}\
  \bibnamefont {Aprile}} \emph {et~al.} (\bibinfo {collaboration} {XENON}),\
  }\href {https://doi.org/10.1103/PhysRevD.103.063028} {\bibfield  {journal}
  {\bibinfo  {journal} {Phys. Rev. D}\ }\textbf {\bibinfo {volume} {103}},\
  \bibinfo {pages} {063028} (\bibinfo {year} {2021})},\ \Eprint
  {https://arxiv.org/abs/2011.10431} {arXiv:2011.10431 [hep-ex]} \BibitemShut
  {NoStop}%
\bibitem [{\citenamefont {Antypas}\ \emph {et~al.}(2022)\citenamefont
  {Antypas}, \citenamefont {Banerjee}, \citenamefont {Bartram}, \citenamefont
  {Baryakhtar}, \citenamefont {Betz}, \citenamefont {Bollinger}, \citenamefont
  {Boutan}, \citenamefont {Bowring}, \citenamefont {Budker}, \citenamefont
  {Carney} \emph {et~al.}}]{antypas_new_2022}%
  \BibitemOpen
  \bibfield  {author} {\bibinfo {author} {\bibfnamefont {D.}~\bibnamefont
  {Antypas}}, \bibinfo {author} {\bibfnamefont {A.}~\bibnamefont {Banerjee}},
  \bibinfo {author} {\bibfnamefont {C.}~\bibnamefont {Bartram}}, \bibinfo
  {author} {\bibfnamefont {M.}~\bibnamefont {Baryakhtar}}, \bibinfo {author}
  {\bibfnamefont {J.}~\bibnamefont {Betz}}, \bibinfo {author} {\bibfnamefont
  {J.~J.}\ \bibnamefont {Bollinger}}, \bibinfo {author} {\bibfnamefont
  {C.}~\bibnamefont {Boutan}}, \bibinfo {author} {\bibfnamefont
  {D.}~\bibnamefont {Bowring}}, \bibinfo {author} {\bibfnamefont
  {D.}~\bibnamefont {Budker}}, \bibinfo {author} {\bibfnamefont
  {D.}~\bibnamefont {Carney}}, \emph {et~al.},\ }\href
  {http://arxiv.org/abs/2203.14915} {\bibfield  {journal} {\bibinfo  {journal}
  {arXiv:2203.14915}\ } (\bibinfo {year} {2022})}\BibitemShut {NoStop}%
\bibitem [{\citenamefont {Damour}\ and\ \citenamefont
  {Donoghue}(2010)}]{damour_phenomenology_2010}%
  \BibitemOpen
  \bibfield  {author} {\bibinfo {author} {\bibfnamefont {T.}~\bibnamefont
  {Damour}}\ and\ \bibinfo {author} {\bibfnamefont {J.}~\bibnamefont
  {Donoghue}},\ }\href {https://doi.org/10.1088/0264-9381/27/20/202001}
  {\bibfield  {journal} {\bibinfo  {journal} {Classical Quantum Gravity}\
  }\textbf {\bibinfo {volume} {27}},\ \bibinfo {pages} {202001} (\bibinfo
  {year} {2010})}\BibitemShut {NoStop}%
\bibitem [{\citenamefont {Pašteka}\ \emph {et~al.}(2019)\citenamefont
  {Pašteka}, \citenamefont {Hao}, \citenamefont {Borschevsky}, \citenamefont
  {Flambaum},\ and\ \citenamefont {Schwerdtfeger}}]{pasteka_material_2019}%
  \BibitemOpen
  \bibfield  {author} {\bibinfo {author} {\bibfnamefont {L.}~\bibnamefont
  {Pašteka}}, \bibinfo {author} {\bibfnamefont {Y.}~\bibnamefont {Hao}},
  \bibinfo {author} {\bibfnamefont {A.}~\bibnamefont {Borschevsky}}, \bibinfo
  {author} {\bibfnamefont {V.}~\bibnamefont {Flambaum}},\ and\ \bibinfo
  {author} {\bibfnamefont {P.}~\bibnamefont {Schwerdtfeger}},\ }\href
  {https://doi.org/10.1103/PhysRevLett.122.160801} {\bibfield  {journal}
  {\bibinfo  {journal} {Phys. Rev. Lett.}\ }\textbf {\bibinfo {volume} {122}},\
  \bibinfo {pages} {160801} (\bibinfo {year} {2019})}\BibitemShut {NoStop}%
\bibitem [{\citenamefont {Marsh}(2016)}]{marsh_axion_2016}%
  \BibitemOpen
  \bibfield  {author} {\bibinfo {author} {\bibfnamefont {D.}~\bibnamefont
  {Marsh}},\ }\href
  {https://doi.org/https://doi.org/10.1016/j.physrep.2016.06.005} {\bibfield
  {journal} {\bibinfo  {journal} {Phys. Rep.}\ }\textbf {\bibinfo {volume}
  {643}},\ \bibinfo {pages} {1} (\bibinfo {year} {2016})},\ \bibinfo {note}
  {axion cosmology}\BibitemShut {NoStop}%
\bibitem [{\citenamefont {Sakharov}\ \emph {et~al.}(1996)\citenamefont
  {Sakharov}, \citenamefont {Sokoloff},\ and\ \citenamefont
  {Khlopov}}]{sakharov1996large}%
  \BibitemOpen
  \bibfield  {author} {\bibinfo {author} {\bibfnamefont {A.}~\bibnamefont
  {Sakharov}}, \bibinfo {author} {\bibfnamefont {D.}~\bibnamefont {Sokoloff}},\
  and\ \bibinfo {author} {\bibfnamefont {M.~Y.}\ \bibnamefont {Khlopov}},\
  }\href@noop {} {\bibfield  {journal} {\bibinfo  {journal} {Physics of Atomic
  Nuclei}\ }\textbf {\bibinfo {volume} {59}},\ \bibinfo {pages} {1005}
  (\bibinfo {year} {1996})}\BibitemShut {NoStop}%
\bibitem [{\citenamefont {Berezhiani}\ and\ \citenamefont
  {Khlopov}(1991)}]{berezhiani1991cosmology}%
  \BibitemOpen
  \bibfield  {author} {\bibinfo {author} {\bibfnamefont {Z.}~\bibnamefont
  {Berezhiani}}\ and\ \bibinfo {author} {\bibfnamefont {M.}~\bibnamefont
  {Khlopov}},\ }\href@noop {} {\bibfield  {journal} {\bibinfo  {journal}
  {Zeitschrift f{\"u}r Physik C Particles and Fields}\ }\textbf {\bibinfo
  {volume} {49}},\ \bibinfo {pages} {73} (\bibinfo {year} {1991})}\BibitemShut
  {NoStop}%
\bibitem [{\citenamefont {Berezhiani}\ \emph {et~al.}(1992)\citenamefont
  {Berezhiani}, \citenamefont {Sakharov},\ and\ \citenamefont
  {Khlopov}}]{berezhiani1992primordial}%
  \BibitemOpen
  \bibfield  {author} {\bibinfo {author} {\bibfnamefont {Z.}~\bibnamefont
  {Berezhiani}}, \bibinfo {author} {\bibfnamefont {A.}~\bibnamefont
  {Sakharov}},\ and\ \bibinfo {author} {\bibfnamefont {M.}~\bibnamefont
  {Khlopov}},\ }\href@noop {} {\bibfield  {journal} {\bibinfo  {journal}
  {Soviet Journal of Nuclear Physics}\ }\textbf {\bibinfo {volume} {55}},\
  \bibinfo {pages} {1063} (\bibinfo {year} {1992})}\BibitemShut {NoStop}%
\bibitem [{\citenamefont {DeRocco}\ and\ \citenamefont
  {Hook}(2018)}]{derocco_axion_2018}%
  \BibitemOpen
  \bibfield  {author} {\bibinfo {author} {\bibfnamefont {W.}~\bibnamefont
  {DeRocco}}\ and\ \bibinfo {author} {\bibfnamefont {A.}~\bibnamefont {Hook}},\
  }\href {https://doi.org/10.1103/PhysRevD.98.035021} {\bibfield  {journal}
  {\bibinfo  {journal} {Phys. Rev. D}\ }\textbf {\bibinfo {volume} {98}},\
  \bibinfo {pages} {035021} (\bibinfo {year} {2018})}\BibitemShut {NoStop}%
\bibitem [{\citenamefont {Nagano}\ \emph {et~al.}(2019)\citenamefont {Nagano},
  \citenamefont {Fujita}, \citenamefont {Michimura},\ and\ \citenamefont
  {Obata}}]{nagano_axion_2019}%
  \BibitemOpen
  \bibfield  {author} {\bibinfo {author} {\bibfnamefont {K.}~\bibnamefont
  {Nagano}}, \bibinfo {author} {\bibfnamefont {T.}~\bibnamefont {Fujita}},
  \bibinfo {author} {\bibfnamefont {Y.}~\bibnamefont {Michimura}},\ and\
  \bibinfo {author} {\bibfnamefont {I.}~\bibnamefont {Obata}},\ }\href
  {https://doi.org/10.1103/PhysRevLett.123.111301} {\bibfield  {journal}
  {\bibinfo  {journal} {Phys. Rev. Lett.}\ }\textbf {\bibinfo {volume} {123}},\
  \bibinfo {pages} {111301} (\bibinfo {year} {2019})}\BibitemShut {NoStop}%
\bibitem [{\citenamefont {Ejlli}\ \emph {et~al.}(2018)\citenamefont {Ejlli},
  \citenamefont {Della~Valle},\ and\ \citenamefont
  {Zavattini}}]{ejlli_polarisation_2018}%
  \BibitemOpen
  \bibfield  {author} {\bibinfo {author} {\bibfnamefont {A.}~\bibnamefont
  {Ejlli}}, \bibinfo {author} {\bibfnamefont {F.}~\bibnamefont {Della~Valle}},\
  and\ \bibinfo {author} {\bibfnamefont {G.}~\bibnamefont {Zavattini}},\ }\href
  {https://doi.org/10.1007/s00340-018-6891-3} {\bibfield  {journal} {\bibinfo
  {journal} {Appl. Phys. B}\ }\textbf {\bibinfo {volume} {124}},\ \bibinfo
  {pages} {22} (\bibinfo {year} {2018})}\BibitemShut {NoStop}%
\bibitem [{\citenamefont {Schwartz}\ \emph {et~al.}(2020)\citenamefont
  {Schwartz}, \citenamefont {Pele}, \citenamefont {Warner}, \citenamefont
  {Lantz}, \citenamefont {Betzwieser}, \citenamefont {Dooley}, \citenamefont
  {Biscans}, \citenamefont {Coughlin}, \citenamefont {Mukund}, \citenamefont
  {Abbott} \emph {et~al.}}]{Schwartz_2020}%
  \BibitemOpen
  \bibfield  {author} {\bibinfo {author} {\bibfnamefont {E.}~\bibnamefont
  {Schwartz}}, \bibinfo {author} {\bibfnamefont {A.}~\bibnamefont {Pele}},
  \bibinfo {author} {\bibfnamefont {J.}~\bibnamefont {Warner}}, \bibinfo
  {author} {\bibfnamefont {B.}~\bibnamefont {Lantz}}, \bibinfo {author}
  {\bibfnamefont {J.}~\bibnamefont {Betzwieser}}, \bibinfo {author}
  {\bibfnamefont {K.}~\bibnamefont {Dooley}}, \bibinfo {author} {\bibfnamefont
  {S.}~\bibnamefont {Biscans}}, \bibinfo {author} {\bibfnamefont
  {M.}~\bibnamefont {Coughlin}}, \bibinfo {author} {\bibfnamefont
  {N.}~\bibnamefont {Mukund}}, \bibinfo {author} {\bibfnamefont
  {R.}~\bibnamefont {Abbott}}, \emph {et~al.},\ }\href
  {https://doi.org/10.1088/1361-6382/abbc8c} {\bibfield  {journal} {\bibinfo
  {journal} {Classical Quantum Gravity}\ }\textbf {\bibinfo {volume} {37}},\
  \bibinfo {pages} {235007} (\bibinfo {year} {2020})}\BibitemShut {NoStop}%
\bibitem [{\citenamefont {Cahillane}\ and\ \citenamefont
  {Mansell}(2022)}]{LIGOreviewpaper}%
  \BibitemOpen
  \bibfield  {author} {\bibinfo {author} {\bibfnamefont {C.}~\bibnamefont
  {Cahillane}}\ and\ \bibinfo {author} {\bibfnamefont {G.}~\bibnamefont
  {Mansell}},\ }\href {https://doi.org/10.3390/galaxies10010036} {\bibfield
  {journal} {\bibinfo  {journal} {Galaxies}\ }\textbf {\bibinfo {volume}
  {10}},\ \bibinfo {pages} {36} (\bibinfo {year} {2022})}\BibitemShut {NoStop}%
\bibitem [{\citenamefont {Satari}\ \emph {et~al.}(2022)\citenamefont {Satari},
  \citenamefont {Blair}, \citenamefont {Ju}, \citenamefont {Blair},
  \citenamefont {Zhao}, \citenamefont {Saygin}, \citenamefont {Meyers},\ and\
  \citenamefont {Lumley}}]{satari_2022}%
  \BibitemOpen
  \bibfield  {author} {\bibinfo {author} {\bibfnamefont {H.}~\bibnamefont
  {Satari}}, \bibinfo {author} {\bibfnamefont {C.}~\bibnamefont {Blair}},
  \bibinfo {author} {\bibfnamefont {L.}~\bibnamefont {Ju}}, \bibinfo {author}
  {\bibfnamefont {D.}~\bibnamefont {Blair}}, \bibinfo {author} {\bibfnamefont
  {C.}~\bibnamefont {Zhao}}, \bibinfo {author} {\bibfnamefont {E.}~\bibnamefont
  {Saygin}}, \bibinfo {author} {\bibfnamefont {P.}~\bibnamefont {Meyers}},\
  and\ \bibinfo {author} {\bibfnamefont {D.}~\bibnamefont {Lumley}},\ }\href
  {https://doi.org/10.1088/1361-6382/ac92b7} {\bibfield  {journal} {\bibinfo
  {journal} {Classical Quantum Gravity}\ }\textbf {\bibinfo {volume} {39}},\
  \bibinfo {pages} {215015} (\bibinfo {year} {2022})}\BibitemShut {NoStop}%
\bibitem [{\citenamefont {Coherent}(2019)}]{mephisto_lasers}%
  \BibitemOpen
  \bibfield  {author} {\bibinfo {author} {\bibnamefont {Coherent}},\ }\href
  {https://content.coherent.com/legacy-assets/pdf/COHR_MephistoNPRO_WP_9_24_19.pdf}
  {\bibinfo {title} {Mephisto lasers -- ultra-low noise and narrow
  linewidth}},\ \bibinfo {howpublished} {White Paper} (\bibinfo {year}
  {2019})\BibitemShut {NoStop}%
\bibitem [{\citenamefont {Rocchi}(2014)}]{Rocchi}%
  \BibitemOpen
  \bibfield  {author} {\bibinfo {author} {\bibfnamefont {A.}~\bibnamefont
  {Rocchi}},\ }\bibinfo {title} {Thermal effects and other wavefront
  aberrations in recycling cavities},\ in\ \href
  {https://doi.org/10.1007/978-3-319-03792-9_9} {\emph {\bibinfo {booktitle}
  {Advanced Interferometers and the Search for Gravitational Waves: Lectures
  from the First VESF School on Advanced Detectors for Gravitational Waves}}}\
  (\bibinfo  {publisher} {Springer, New York},\ \bibinfo {year} {2014})\ pp.\
  \bibinfo {pages} {251--274}\BibitemShut {NoStop}%
\bibitem [{\citenamefont {Aiello}(2019)}]{AielloPhD}%
  \BibitemOpen
  \bibfield  {author} {\bibinfo {author} {\bibfnamefont {L.}~\bibnamefont
  {Aiello}},\ }\emph {\bibinfo {title} {Development of new approaches for
  optical aberration control in gravitational wave interferometers}},\ \href
  {https://iris.gssi.it/handle/20.500.12571/9702} {Ph.D. thesis},\ \bibinfo
  {school} {SISSA (Scuola Internazionale di Studi Superiori Avanzati) - GSSI
  (Gran Sasso Science Institute)} (\bibinfo {year} {2019})\BibitemShut
  {NoStop}%
\bibitem [{\citenamefont {P.}\ and\ \citenamefont {Vinet}(1990)}]{Hello:Vinet}%
  \BibitemOpen
  \bibfield  {author} {\bibinfo {author} {\bibfnamefont {H.}~\bibnamefont
  {P.}}\ and\ \bibinfo {author} {\bibfnamefont {J.}~\bibnamefont {Vinet}},\
  }\href {https://doi.org/10.1051/jphys:0199000510120126700} {\bibfield
  {journal} {\bibinfo  {journal} {J. Phys.}\ }\textbf {\bibinfo {volume}
  {51}},\ \bibinfo {pages} {1377} (\bibinfo {year} {1990})}\BibitemShut
  {NoStop}%
\bibitem [{Fin(2022)}]{Finesse3}%
  \BibitemOpen
  \href {https://finesse.docs.ligo.org/finesse3/index.html} {\bibinfo {title}
  {Frequency domain interferometer simulation software, version 3}} (\bibinfo
  {year} {2022})\BibitemShut {NoStop}%
\bibitem [{\citenamefont {Hello}(2001)}]{Hello_2001}%
  \BibitemOpen
  \bibfield  {author} {\bibinfo {author} {\bibfnamefont {P.}~\bibnamefont
  {Hello}},\ }\href {https://doi.org/10.1007/s100530170154} {\bibfield
  {journal} {\bibinfo  {journal} {Eur. Phys. J. D}\ }\textbf {\bibinfo {volume}
  {15}},\ \bibinfo {pages} {373} (\bibinfo {year} {2001})}\BibitemShut
  {NoStop}%
\bibitem [{\citenamefont {Aiello}\ \emph {et~al.}(2019)\citenamefont {Aiello},
  \citenamefont {Cesarini}, \citenamefont {Fafone}, \citenamefont {Lorenzini},
  \citenamefont {Minenkov}, \citenamefont {Nardecchia}, \citenamefont
  {Rocchi},\ and\ \citenamefont {Sequino}}]{Aiello_2019}%
  \BibitemOpen
  \bibfield  {author} {\bibinfo {author} {\bibfnamefont {L.}~\bibnamefont
  {Aiello}}, \bibinfo {author} {\bibfnamefont {E.}~\bibnamefont {Cesarini}},
  \bibinfo {author} {\bibfnamefont {V.}~\bibnamefont {Fafone}}, \bibinfo
  {author} {\bibfnamefont {M.}~\bibnamefont {Lorenzini}}, \bibinfo {author}
  {\bibfnamefont {Y.}~\bibnamefont {Minenkov}}, \bibinfo {author}
  {\bibfnamefont {I.}~\bibnamefont {Nardecchia}}, \bibinfo {author}
  {\bibfnamefont {A.}~\bibnamefont {Rocchi}},\ and\ \bibinfo {author}
  {\bibfnamefont {V.}~\bibnamefont {Sequino}},\ }\href
  {https://doi.org/10.1088/1742-6596/1226/1/012019} {\bibfield  {journal}
  {\bibinfo  {journal} {J. Phys. Conf. Ser.}\ }\textbf {\bibinfo {volume}
  {1226}},\ \bibinfo {pages} {012019} (\bibinfo {year} {2019})}\BibitemShut
  {NoStop}%
\bibitem [{\citenamefont {Marchi\`o}\ \emph {et~al.}(2021)\citenamefont
  {Marchi\`o}, \citenamefont {Leonardi}, \citenamefont {Bazzan},\ and\
  \citenamefont {Flaminio}}]{Marchio:2021xco}%
  \BibitemOpen
  \bibfield  {author} {\bibinfo {author} {\bibfnamefont {M.}~\bibnamefont
  {Marchi\`o}}, \bibinfo {author} {\bibfnamefont {M.}~\bibnamefont {Leonardi}},
  \bibinfo {author} {\bibfnamefont {M.}~\bibnamefont {Bazzan}},\ and\ \bibinfo
  {author} {\bibfnamefont {R.}~\bibnamefont {Flaminio}},\ }\href
  {https://doi.org/10.1038/s41598-020-80313-1} {\bibfield  {journal} {\bibinfo
  {journal} {Sci. Rep.}\ }\textbf {\bibinfo {volume} {11}},\ \bibinfo {pages}
  {2654} (\bibinfo {year} {2021})}\BibitemShut {NoStop}%
\bibitem [{\citenamefont {Della~Valle}\ \emph {et~al.}(2014)\citenamefont
  {Della~Valle}, \citenamefont {Ejlli}, \citenamefont {Gastaldi}, \citenamefont
  {Messineo}, \citenamefont {Milotti}, \citenamefont {Pengo}, \citenamefont
  {Piemontese}, \citenamefont {Ruoso},\ and\ \citenamefont
  {Zavattini}}]{della_valle_measurement_2014}%
  \BibitemOpen
  \bibfield  {author} {\bibinfo {author} {\bibfnamefont {F.}~\bibnamefont
  {Della~Valle}}, \bibinfo {author} {\bibfnamefont {A.}~\bibnamefont {Ejlli}},
  \bibinfo {author} {\bibfnamefont {U.}~\bibnamefont {Gastaldi}}, \bibinfo
  {author} {\bibfnamefont {G.}~\bibnamefont {Messineo}}, \bibinfo {author}
  {\bibfnamefont {E.}~\bibnamefont {Milotti}}, \bibinfo {author} {\bibfnamefont
  {R.}~\bibnamefont {Pengo}}, \bibinfo {author} {\bibfnamefont
  {L.}~\bibnamefont {Piemontese}}, \bibinfo {author} {\bibfnamefont
  {G.}~\bibnamefont {Ruoso}},\ and\ \bibinfo {author} {\bibfnamefont
  {G.}~\bibnamefont {Zavattini}},\ }\href
  {https://doi.org/10.1016/j.cplett.2013.12.049} {\bibfield  {journal}
  {\bibinfo  {journal} {Chem. Phys. Lett.}\ }\textbf {\bibinfo {volume}
  {592}},\ \bibinfo {pages} {288} (\bibinfo {year} {2014})}\BibitemShut
  {NoStop}%
\bibitem [{\citenamefont {Savalle}\ \emph {et~al.}(2021)\citenamefont
  {Savalle}, \citenamefont {Hees}, \citenamefont {Frank}, \citenamefont
  {Cantin}, \citenamefont {Pottie}, \citenamefont {Roberts}, \citenamefont
  {Cros}, \citenamefont {McAllister},\ and\ \citenamefont
  {Wolf}}]{savalle_searching_2021}%
  \BibitemOpen
  \bibfield  {author} {\bibinfo {author} {\bibfnamefont {E.}~\bibnamefont
  {Savalle}}, \bibinfo {author} {\bibfnamefont {A.}~\bibnamefont {Hees}},
  \bibinfo {author} {\bibfnamefont {F.}~\bibnamefont {Frank}}, \bibinfo
  {author} {\bibfnamefont {E.}~\bibnamefont {Cantin}}, \bibinfo {author}
  {\bibfnamefont {P.}~\bibnamefont {Pottie}}, \bibinfo {author} {\bibfnamefont
  {B.}~\bibnamefont {Roberts}}, \bibinfo {author} {\bibfnamefont
  {L.}~\bibnamefont {Cros}}, \bibinfo {author} {\bibfnamefont {B.}~\bibnamefont
  {McAllister}},\ and\ \bibinfo {author} {\bibfnamefont {P.}~\bibnamefont
  {Wolf}},\ }\href {https://doi.org/10.1103/PhysRevLett.126.051301} {\bibfield
  {journal} {\bibinfo  {journal} {Phys. Rev. Lett.}\ }\textbf {\bibinfo
  {volume} {126}},\ \bibinfo {pages} {051301} (\bibinfo {year}
  {2021})}\BibitemShut {NoStop}%
\bibitem [{\citenamefont {Tretiak}\ \emph {et~al.}(2022)\citenamefont
  {Tretiak}, \citenamefont {Zhang}, \citenamefont {Figueroa}, \citenamefont
  {Antypas}, \citenamefont {Brogna}, \citenamefont {Banerjee}, \citenamefont
  {Perez},\ and\ \citenamefont {Budker}}]{tretiak_improved_2022}%
  \BibitemOpen
  \bibfield  {author} {\bibinfo {author} {\bibfnamefont {O.}~\bibnamefont
  {Tretiak}}, \bibinfo {author} {\bibfnamefont {X.}~\bibnamefont {Zhang}},
  \bibinfo {author} {\bibfnamefont {N.~L.}\ \bibnamefont {Figueroa}}, \bibinfo
  {author} {\bibfnamefont {D.}~\bibnamefont {Antypas}}, \bibinfo {author}
  {\bibfnamefont {A.}~\bibnamefont {Brogna}}, \bibinfo {author} {\bibfnamefont
  {A.}~\bibnamefont {Banerjee}}, \bibinfo {author} {\bibfnamefont
  {G.}~\bibnamefont {Perez}},\ and\ \bibinfo {author} {\bibfnamefont
  {D.}~\bibnamefont {Budker}},\ }\href
  {https://doi.org/10.1103/PhysRevLett.129.031301} {\bibfield  {journal}
  {\bibinfo  {journal} {Phys. Rev. Lett.}\ }\textbf {\bibinfo {volume} {129}},\
  \bibinfo {pages} {031301} (\bibinfo {year} {2022})}\BibitemShut {NoStop}%
\bibitem [{\citenamefont {Anastassopoulos}\ \emph {et~al.}(2017)\citenamefont
  {Anastassopoulos}, \citenamefont {Aune}, \citenamefont {Barth}, \citenamefont
  {Belov}, \citenamefont {Cantatore}, \citenamefont {Carmona}, \citenamefont
  {Castel}, \citenamefont {Cetin}, \citenamefont {Christensen}, \citenamefont
  {Collar} \emph {et~al.}}]{anastassopoulos2017new}%
  \BibitemOpen
  \bibfield  {author} {\bibinfo {author} {\bibfnamefont {V.}~\bibnamefont
  {Anastassopoulos}}, \bibinfo {author} {\bibfnamefont {S.}~\bibnamefont
  {Aune}}, \bibinfo {author} {\bibfnamefont {K.}~\bibnamefont {Barth}},
  \bibinfo {author} {\bibfnamefont {A.}~\bibnamefont {Belov}}, \bibinfo
  {author} {\bibfnamefont {G.}~\bibnamefont {Cantatore}}, \bibinfo {author}
  {\bibfnamefont {J.}~\bibnamefont {Carmona}}, \bibinfo {author} {\bibfnamefont
  {J.}~\bibnamefont {Castel}}, \bibinfo {author} {\bibfnamefont
  {S.}~\bibnamefont {Cetin}}, \bibinfo {author} {\bibfnamefont
  {F.}~\bibnamefont {Christensen}}, \bibinfo {author} {\bibfnamefont
  {J.}~\bibnamefont {Collar}}, \emph {et~al.},\ }\href
  {https://doi.org/10.1038/nphys4109} {\bibfield  {journal} {\bibinfo
  {journal} {Nat. Physics}\ }\textbf {\bibinfo {volume} {13}},\ \bibinfo
  {pages} {584–590} (\bibinfo {year} {2017})}\BibitemShut {NoStop}%
\bibitem [{\citenamefont {Bähre}\ \emph {et~al.}(2013)\citenamefont {Bähre},
  \citenamefont {Döbrich}, \citenamefont {Dreyling-Eschweiler}, \citenamefont
  {Ghazaryan}, \citenamefont {Hodajerdi}, \citenamefont {Horns}, \citenamefont
  {Januschek}, \citenamefont {Knabbe}, \citenamefont {Lindner}, \citenamefont
  {Notz} \emph {et~al.}}]{Bahre:2013ywa}%
  \BibitemOpen
  \bibfield  {author} {\bibinfo {author} {\bibfnamefont {R.}~\bibnamefont
  {Bähre}}, \bibinfo {author} {\bibfnamefont {B.}~\bibnamefont {Döbrich}},
  \bibinfo {author} {\bibfnamefont {J.}~\bibnamefont {Dreyling-Eschweiler}},
  \bibinfo {author} {\bibfnamefont {S.}~\bibnamefont {Ghazaryan}}, \bibinfo
  {author} {\bibfnamefont {R.}~\bibnamefont {Hodajerdi}}, \bibinfo {author}
  {\bibfnamefont {D.}~\bibnamefont {Horns}}, \bibinfo {author} {\bibfnamefont
  {F.}~\bibnamefont {Januschek}}, \bibinfo {author} {\bibfnamefont
  {E.}~\bibnamefont {Knabbe}}, \bibinfo {author} {\bibfnamefont
  {A.}~\bibnamefont {Lindner}}, \bibinfo {author} {\bibfnamefont
  {D.}~\bibnamefont {Notz}}, \emph {et~al.},\ }\href
  {https://doi.org/10.1088/1748-0221/8/09/T09001} {\bibfield  {journal}
  {\bibinfo  {journal} {J. Instrum.}\ }\textbf {\bibinfo {volume} {8}},\
  \bibinfo {pages} {T09001}}\BibitemShut {NoStop}%
\bibitem [{\citenamefont {Punturo}\ \emph {et~al.}(2010)\citenamefont
  {Punturo}, \citenamefont {Abernathy}, \citenamefont {Acernese}, \citenamefont
  {Allen}, \citenamefont {Andersson}, \citenamefont {Arun}, \citenamefont
  {Barone}, \citenamefont {Barr}, \citenamefont {Barsuglia}, \citenamefont
  {Beker} \emph {et~al.}}]{Punturo_2010}%
  \BibitemOpen
  \bibfield  {author} {\bibinfo {author} {\bibfnamefont {M.}~\bibnamefont
  {Punturo}}, \bibinfo {author} {\bibfnamefont {M.}~\bibnamefont {Abernathy}},
  \bibinfo {author} {\bibfnamefont {F.}~\bibnamefont {Acernese}}, \bibinfo
  {author} {\bibfnamefont {B.}~\bibnamefont {Allen}}, \bibinfo {author}
  {\bibfnamefont {N.}~\bibnamefont {Andersson}}, \bibinfo {author}
  {\bibfnamefont {K.}~\bibnamefont {Arun}}, \bibinfo {author} {\bibfnamefont
  {F.}~\bibnamefont {Barone}}, \bibinfo {author} {\bibfnamefont
  {B.}~\bibnamefont {Barr}}, \bibinfo {author} {\bibfnamefont {M.}~\bibnamefont
  {Barsuglia}}, \bibinfo {author} {\bibfnamefont {M.}~\bibnamefont {Beker}},
  \emph {et~al.},\ }\href {https://doi.org/10.1088/0264-9381/27/19/194002}
  {\bibfield  {journal} {\bibinfo  {journal} {Classical Quantum Gravity}\
  }\textbf {\bibinfo {volume} {27}},\ \bibinfo {pages} {194002} (\bibinfo
  {year} {2010})}\BibitemShut {NoStop}%
\bibitem [{\citenamefont {\textit{et al} (LIGO
  Scientific~Collaboration)}(2017)}]{Abbott_2017}%
  \BibitemOpen
  \bibfield  {author} {\bibinfo {author} {\bibfnamefont {B.~P.~A.}\
  \bibnamefont {\textit{et al} (LIGO Scientific~Collaboration)}},\ }\href
  {https://doi.org/10.1088/1361-6382/aa51f4} {\bibfield  {journal} {\bibinfo
  {journal} {Classical Quantum Gravity}\ }\textbf {\bibinfo {volume} {34}},\
  \bibinfo {pages} {044001} (\bibinfo {year} {2017})}\BibitemShut {NoStop}%
\end{thebibliography}%


%

\end{document}